\documentclass[11pt]{article}

\usepackage{amssymb,amsmath,amsfonts,eurosym,geometry,ulem,graphicx,caption,color,setspace,sectsty,comment,footmisc,caption,pdflscape,subfigure,array,lscape,booktabs, epigraph, xfrac,placeins, array, titlesec}

\usepackage[hidelinks]{hyperref}
\usepackage{mathptmx}

\usepackage[labelfont=bf]{caption}
\usepackage[colorinlistoftodos]{todonotes}

\usepackage{natbib}
\usepackage{afterpage}

\titlespacing\section{0pt}{10pt plus 4pt minus 2pt}{0pt plus 2pt minus 2pt}
\titlespacing\subsection{0pt}{10pt plus 4pt minus 2pt}{0pt plus 2pt minus 2pt}
\titlespacing\subsubsection{0pt}{10pt plus 4pt minus 2pt}{0pt plus 2pt minus 2pt}

\usepackage{authblk}

\begin{document}

\begin{titlepage}
\title{A narrowing of AI research?}

\author[1]{Joel Klinger}
\author[2]{Juan Mateos-Garcia}
\author[3]{Konstantinos Statholoupoulos}

\affil[1,2]{Nesta, 58 Victoria Embankment EC4Y 0DS}
\affil[3]{Util, 8 Hermitage St. Paddington, London W2 1BE}

\date{\today}
\maketitle
\begin{abstract}
\singlespacing
\noindent
The arrival of deep learning techniques able to infer patterns from large datasets has
dramatically improved the performance of Artificial Intelligence (AI) systems. Deep learning's rapid development and adoption, in great part led by large technology companies, has however created concerns about a premature narrowing in the technological trajectory of AI research despite its weaknesses, which include lack of robustness, high environmental costs, and potentially unfair outcomes. We seek to improve the evidence base with a semantic analysis of AI research in arXiv, a popular pre-prints database. We study the evolution of the thematic diversity of AI research, compare the thematic diversity of AI research in academia and the private sector and measure the influence of private companies in AI research through the citations they receive and their collaborations with other institutions. Our results suggest that diversity in AI research has stagnated in recent years, and that AI research involving the private sector tends to be less diverse and more influential than research in academia. We also find that private sector AI researchers tend to specialise in data-hungry and computationally intensive deep learning methods at the expense of research involving other AI methods, research that considers the societal and ethical implications of AI, and applications in sectors like health. Our results provide a rationale for policy action to prevent a premature narrowing of AI research that could constrain its societal benefits, but we note the informational, incentive and scale hurdles standing in the way of such interventions.

\end{abstract}
\setcounter{page}{0}
\thispagestyle{empty}
\end{titlepage}
\pagebreak \newpage

\singlespacing

\epigraph{\textit{Tenembaum told me he looks to the story of backprop for inspiration. For decades, backprop was cool math that didn’t really accomplish anything. As computers got faster and the engineering got more sophisticated, suddenly it did. He hopes the same thing might happen with his own work and that of his students, ``but it might take another couple decades''}.}{Interview with AI researcher Joshua Tenenbaum in \citet{somersis2017}}

\section{Introduction}\label{sec:introduction}
Technological change has a direction as well as a rate: different designs for a technology are possible, and some of them may be more societally desirable than others \citep{aghion2009science}. Fortuitous events, shortsightedness and lack of coordination can however create situations where an inferior design becomes dominant and hard to switch away from even after its limitations become apparent \citep{david1985clio, arthur1994increasing}. Consider for example the automobile, where the combustion engine surpassed environmentally friendlier alternatives based on steam or electricity, or the case of the nuclear reactor, where early demand from the military locked this technology into a light water design that was less suitable for civilian applications \citep{cowan_1990}. 

Faced with uncertainty about the benefits and risks of alternative technologies and the desire to avoid premature lock-in to an inferior design, it might be reasonable to put in place policies to preserve diversity in the technology landscape \citep{aghion2009science}. 
% In military procurement, funders have long adopted portfolio strategies to explore multiple technologies in parallel \citep{johnsonb}. Contemporary approaches to mission-oriented innovation policy call for bottom-up exploration of technological opportunities that avoid picking a single solution to tackle complex societal challenges \citep{mazzucato2018mission}. 

% who have respectively heralded AI as the latest example of a General Purpose Technology with revolutionary potential  and launched national strategies and plans to bolster indigenous AI industries and mitigate the disruption caused by labour-displacing AI systems \citep{paunov}.

Are such initiatives needed to sustain technological diversity in Artificial Intelligence (AI) research? AI systems based on deep learning, a machine learning technique that infers patterns from large unstructured datasets have been deployed successfully in many digital and media products and services \citep{lecun2015deep}. AI's technical and commercial successes have attracted large R\&D investments from corporations and venture capitalists and the attention of economists and policymakers \citep{cockburn2018impact, paunov}. Others have highlighted the limitations of this AI trajectory in terms of the brittleness of the systems it creates, its environmental costs and the risk of discrimination when data-hungry AI systems are trained on biased data \citep{marcus2018deep, damour2020underspecification, thompson2020computational, strubell2020energy, buolamwini2018gender, bender2021dangers}.

% Some researchers note that AI systems based on deep learning are brittle and liable to under-perform or even fail in unexpected and catastrophic ways when they are exposed to new situations or gamed by malicious actors, potentially making them unsuitable for high-stake domains such as transport or health \citep{marcus2018deep, russell2019,damour2020underspecification}. Economists worry that businesses may deploy `mediocre' AI systems that fail to offset their labour displacement effects with productivity gains \citep{acemogluwrong2019, acemoglu2021harms}. Activists, journalists and critical scholars of technology have found that some AI systems generate discriminatory and unfair outcomes \citep{buolamwini2018gender} and/or enable mass surveillance by governments and commercial actors \citep{zuboff2019age} . There is increasing evidence that data-hungry, computationally intensive deep learning systems may be environmentally unsustainable \citep{thompson2020computational, strubell2020energy}. 

The terms of the debate echo a three-step process often described in the path dependency literature: a powerful yet in some ways flawed technological design gains momentum among researchers, entrepreneurs, investors and policymakers (step 1), reducing the diversity of ideas that are explored and creating the risk of lock-in (step 2) that may be regretted in the future, when the full costs of the dominant design become manifest (step 3).  In this paper we study the first two steps of this process in the context of AI research: we analyse the evolution of its thematic diversity paying particular attention to the activities of private sector organisations that have become increasingly important in AI research \citep{hain2020, wahmed2020,hagendorff2020ethical, whittaker2021steep} and which, the literature suggests, could make it narrower by concentrating on those technology designs with the shortest-term commercial potential regardless of their longer term impacts and externalities \citep{bryan2017direction}.\footnote{Throughout the paper we will use the term thematic diversity to refer to the topics covered in our corpus - they mostly comprise research and technological subjects so our usage of the term is closely related to notions of research and technological diversity.} 

To do this, we analyse the corpus of AI research in arXiv, a pre-prints repository widely used by the AI research and development community. We estimate the thematic composition of this corpus with a topic model and use this information to calculate its thematic diversity according to various metrics. We study the evolution of these metrics, compare the diversity of AI research involving private firms with the rest of the corpus and analyse the influence of private sector AI research through its citations and collaborations.

% Our results suggest that thematic diversity in AI research has stagnated and even declined in recent years, and that private sector organisations are more narrowly focused than organisations in academia and the public sector on frontier, computationally demanding deep learning technologies. Our results suggest that there may be a policy rationale to preserve diversity in AI research. The metrics that we develop could help inform this agenda.

In Section \ref{sec:literature} we review relevant ideas from the literature which we set in the history of AI and recent debates about its trajectory. In Section \ref{sec:data} we describe our research questions, data and methods and in Section \ref{sec:result} we present our findings. Section \ref{sec:conclusion} concludes with a discussion of our results, their limitations, issues for further research and policy implications.

\section{Literature Review}
\label{sec:literature}

In this section, we review the relevant literatures with the goal of formulating four claims: that multiple technologies fulfilling the same purpose can co-exist; that this  technological diversity may be desirable in a context of uncertainty about the benefits and risks of different technological designs; that there are important economic forces that militate against technological diversity, and that private sector firms in particular may have less incentives to preserve technological diversity than other actors in the research and innovation ecosystem. We connect each of these ideas to defining features of AI, its history and recent debates about its dominant trajectory and the influential role that large technology companies are playing in its development (see Table \ref{tab1_lit_ai_connections} for a summary).

\begin{table}[!h]
    \centering
    \small
    \begin{tabular}{p{2.7in} p{2.7in}}
    \toprule
    Claim & AI manifestation  \\
    \midrule
Many alternative technology designs are possible & Various technology designs explored in different eras of AI research \\
\\Technological diversity is valuable in the face of uncertainty & Concerns about deep learning's safety, environmental and ethical risks \\
\\ Network effects and social dilemmas militate against technological diversity & Hardware lotteries and races between actors create the risk of premature lock-in and shortsighted decisions \\
\\ Private companies have less incentives to preserve technological diversity than other actors in the research and innovation system & Perception of a de-democratization of AI research and private sector dominance over research agendas \\
    \bottomrule
    \end{tabular}
    \caption{Claims based on the literature and their manifestation in the history and current situation of AI R\& D}
    \label{tab1_lit_ai_connections}
\end{table}

\subsection{Technological directionality and AI's technological trajectory}
\label{subsec:direction}

Following Arthur, we define a technology as \textit{`a means to fulfil a human purpose'} and \textit{`an assemblage of practices and components'} \citep{arthur2009nature}. The notion of directionality in technological change reflects the fact that there are generally many different components and assemblages of components (which we will refer to as `technological designs') that could be feasibly deployed to build a technology that fulfils a human purpose: for example, an automobile can be based on a combustion engine design, a steam-powered design or an electricity-powered design.

% This has two implications: first, and in principle, different designs of the same technology could co-exist, fulfilling the same purpose through different combinations of components; second, in the face of such diversity, technological evolution will rarely be a deterministic process:  given a set of technological designs that could fulfil a human purpose and a set of contexts where those designs could be deployed, we would not expect to find a design such that its performance is superior in all contexts. If such design existed, then it would always be adopted regardless of the context and it would be unnecessary to consider its alternatives. The design would be equivalent to the technology, and its direction would be singular.\footnote{We could characterise it with a single value, its rate.} If to the contrary the relative performance of a technological design depends on its context, this means that multiple technological designs are in principle possible, and the direction of a technology matters. 

When applying these ideas to AI technological systems \textit{`that receive percepts from the environment and take actions that affect that environment'} \citep{russell_norvig}, we conceptualise these systems as a combination of data, software, computational infrastructure and hardware \citep{brundage2016modeling, vannuccini2021artificial}. These systems are organised according to architectures or designs that determine what kind of data can be used as an input into an AI system, the computational costs of deploying the system, the characteristics of its outputs, and the opportunities to incorporate them in various products and services, thus shaping its trajectory of development \citep{dosi1982technological, henderson1990architectural}. 

Consistent with the idea of directionality, AI researchers and technologists have explored various designs to build AI systems through the history of the field. In the 1950s they used symbolic methods to implement logical behaviours in computers, and in the 1980s they created Expert Systems embedding the knowledge and heuristics of human experts \citep{wooldridge2020}. More recently, they have adopted machine learning algorithms that are trained to infer patterns from labelled datasets. The 2010s in particular saw important advances in deep learning \citep{lecun2015deep}, a machine learning technique loosely inspired by the operation of the human brain where artificial neurons learn abstract patterns from large, unstructured datasets using backpropagation, an algorithm that adjusts the strength of the links between neurons in the network to minimise prediction error \citep{chauvin1995backpropagation}. Deep learning has contributed to rapid advances in a variety of AI tasks including computer vision and image and video generation, speech recognition and synthesis, translation, question answering, robotics and game playing, and has been successfully deployed in mainstream products and services such as search engines, social networking sites, translation systems, digital personal assistants and self-driving vehicles \citep{agrawal2018prediction}. 

% According to the AI Index, a project to measure various dimensions of the AI ecosystem, global private investment in AI in 2019 amounted to \$70Bn. Funding for AI startups has been increasing at an average annual growth rate of 48\% between 2010 and 2018. The OECD AI Observatory has identified over 300 policy initiatives related to AI in sixty countries. 

\subsection{The value of technological diversity}
\label{subsec:diversity_value}

The notion of technological diversity captures the degree to which a heterogeneous set of technologies addressing similar purposes co-exist in a market or sector.\footnote{We provide a formal operationalisation of this definition in Section \ref{sec:data}.}. There are three reasons why technological diversity may be a desirable feature of a market, particularly in conditions of uncertainty about the benefits and costs of different technological designs and the trajectories that they support.

The first is \emph{resilience}: homogeneous technological ecosystems are more vulnerable to changes in circumstances including the discovery of unexpected defects or limitations in a dominant design in the absence of alternative technologies to switch into \citep{acemoglu2011diversity}. For example, the depletion of global oil reserves and the recognition of the environmental impact of \(CO_{2}\) emissions have called into question society's reliance on combustion engines. 

Along similar lines, there is increasing evidence of important weaknesses in AI systems based on deep learning: these systems have limited ability to generalise outside of the datasets they were originally trained on \citep{damour2020underspecification, marcus2018deep}, which makes them unsafe when exposed to novel situations or when confronted by malicious users \citep{brundage2018malicious}. Deep learning systems require large-scale computational infrastructures for training and deployment, creating a substantial carbon footprint \citep{thompson2020computational, strubell2020energy}. Lastly, deep learning systems perform better when they are trained in very large datasets, with the downside that these datasets are often too large to be curated / filtered to remove personal information and discriminatory and/or inflammatory content. This creates privacy issues and the risk that the resulting models will be biased and unfair towards minorities and disadvantaged groups \citep{raji2021ai, bender2021dangers}.

% \begin{itemize}
% \item \textit{Brittleness:} deep learning systems have limited ability to generalise outside of the datasets they were originally trained on. This makes them unsafe when exposed to novel situations or when confronted by malicious users who may, for example, use `adversarial examples' (subtly modified inputs) to induce erratic and nonsensical behaviours in deep learning systems. 
% \item \textit{Opacity:} it is difficult to interpret the procedures through which a deep learning system transforms inputs into predictions or recommendations. 
% \item \textit{Environmental costs:} Deep learning systems require large-scale computational infrastructures for training and deployment, creating a substantial carbon footprint. For example, it has been estimated that training a single state-of-the-art transformer model in 2019 produced the same carbon emissions as the lifetime of six cars. cite-strubell
% \item \textit{Ethical risks:} Deep learning systems perform better when they are trained in very large datasets (this is in fact one of the strategies to reduce their brittleness). One downside of this is that the data inputs for state-of-the-art AI systems are frequently too large to be curated / filtered to remove personal information and discriminatory and/or inflammatory content, creating privacy issues and the risk that the resulting models will be biased and unfair towards minorities and disadvantaged groups. cite-raji, cite-bender
% \end{itemize}

A second reason to preserve technological diversity is \emph{inclusion}: some technologies may need to be deployed in a wide range of circumstances, making it necessary to maintain multiple designs so as to avoid excluding specific user groups and communities from their benefits \citep{stirling2011pluralising}. There are also political / ethical reasons that make technological diversity desirable: if specific technology designs are aligned with particular political philosophies or values \citep{winner1980artifacts, gabriel2020artificial}, then this means that preserving technological diversity could be in the interest of political pluralism and inter-generational solidarity.

In the case of AI, some sectors such as advertising and media benefit from an abundance of data that can be used to train and target deep learning models. Other sectors, such as education, are less data-intensive. The high-stakes nature of health and medical decisions renders opaque deep learning less-suitable for that sector than in social media or search applications \citep{marcus2019rebooting}. A loss of technological diversity in AI could lead to a situation where some sectors or communities lack AI systems adapted to their needs.

A third and final reason to preserve technological diversity is \emph{creativity}: innovation involves the creative recombination of ideas, and unusual mixes are often an important source of radical and transformative innovations (Arthur 2009). For example, today's deep learning methods emerged at the intersection of computer science and neuroscience \citep{lecun2015deep}. Some recent advances such as the AlphaGo program that defeated Go world champion Lee Sedol in 2016 also brought together state-of-the-art deep reinforcement learning techniques and traditional tree-search algorithms \citep{pumperla2019deep}. Many researchers believe that some of deep learning's limitations could be overcome with techniques from other AI traditions such as symbolic logic, causal inference or intelligence augmentation that have been sidelined during the current `AI boom' (Marcus and Davis 2019; Pearl 2018). A homogeneous landscape of AI research would contain a less varied set of ideas that can be recombined in this way.

\subsection{Drivers of technological homogenisation}
\label{subsec:driver_homog}

High levels of technological diversity seem to be the exception rather than the norm in most markets: until recently, almost all automobiles were based on the combustion engine, and notwithstanding its limitations, deep learning appears to be becoming the dominant method to build state-of-the-art AI systems. What processes drive this loss of technological diversity?

Here, we will focus on two mutually reinforcing processes under two broad rubrics: economies of scale, and social dilemmas. 

\emph{Economies of scale} can narrow technological diversity in several ways. First, there are \textit{supply-side} economies of scale: the development of any new technology requires a substantial investment in knowledge creation, which is characterised by high fixed costs to create the first copy or blueprint, and low marginal costs to produce subsequent versions \citep{arrow2015economic}. This is reflected in product life-cycles where, after an initial phase of exploration of alternative technological options, a dominant design emerges and the locus of competition shifts from product innovation to process innovation that favours larger organizations able to reap economies of scale in the deployment of the design \citep{utterback1975dynamic, abernathy1978patterns, suarez1995dominant}. Second, there are \textit{demand-side} economies of scale if/when a technological design becomes more attractive the more widely adopted it is. Third, there are \textit{two-sided market} economies of scale when the adoption of a technological design creates incentives to invest in complementary resources such as skills, infrastructure or compatible technologies whose availability makes that design more attractive in a positive feedback loop \citep{rochet2006two}. These `network effects' create the risk that fortuitous events early in the trajectory of a technology might result in lock-in to a design irrespective of its merits compared to the alternatives, and a winner-takes-all outcome where a single design dominates the technology landscape \citep{arthur1994increasing}. The case of the QWERTY keyboard is a paradigmatic example of this process \cite{david1985clio}. 

AI's history highlights the importance of network effects and serendipity for its technological trajectory. \citet{hooker2020hardware} argues that the lag between the development of key AI designs and their deployment is an example of `hardware lotteries' in AI research: this means that the adoption of an idea depends not only on its merits but also on the availability of suitable complements such as hardware, software and data to implement it. For example, many of the key concepts underlying deep learning had been introduced in the 1950s and 1980s but it was not until the 2010s, with the increasing availability of suitable hardware, software and data, that these methods could be implemented at scale \citep{wooldridge2020}. One hardware innovation in particular - Graphics Processor Units (GPUs) that had been developed independently for video-games applications - became an important enabler for deep learning techniques. Once deep learning `won' this hardware lottery, it started attracting large investments that sped up its development, leading to the collection of additional data and creation of computational infrastructure in a positive feedback loop that strengthens its position.

\emph{Social dilemmas} are brought about by the public good aspects of technological diversity: a single actor is unlikely to fully capture the benefits of her investments on designs that preserve it, making such investments less likely.

\cite{acemoglu2011diversity} studies these processes with a model where researchers choose between developing a dominant technology or preserving a second-tier alternative, showing that they have strong incentives to contribute to the dominant technology even though preserving technological diversity would be desirable. \cite{hopenhayn} show that competitive allocations of R\&D resources create a social dilemma with over-investment and cannibalisation of efforts in those areas that are perceived to be high return and insufficient exploration of other areas where problems remain unsolved. \citet{bryan2017direction} consider a scenario with sequential research where discoveries build on each other. In this model, the laissez-faire outcome is directionally inefficient because firms race in pursuit of those research lines that are most beneficial in the short-term at the expense of alternatives with larger long-term benefits.

In the case of AI, there is a growing perception that publication, commercial and geopolitical races could be encouraging these shortsighted behaviours (Armstrong, Bostrom, and Shulman 2016). AI development and deployment are frequently described as `winner-takes-all' processes: private sector companies that lead on AI will be able to dominate their markets and expand into new ones. The countries that control the direction of AI development will be able to assert their political systems and values \citep{lee_ai_2018}. This narrative is leading to races in AI development and deployment \citep{armstrong2016racing}. Some researchers have raised concerns that this could result in a `race to the bottom' in AI safety, and that the pressure to achieve state-of-the-art results is creating `troubling trends' in AI scholarship \citep{Lipton_2019}. This sense of urgency may also lead them to focus on technologies that generate short-term returns, even if they are aware that there are alternatives that would be more beneficial in the longer term.

\subsection{The private sector and AI's technological diversity}
\label{subsec:lit_private}

We would expect private sector involvement in the development of a technology to intensify the diversity-narrowing processes we described above: commercial actors have strong incentives to leverage their technology investments across more markets, and to focus on those technologies that can be more readily deployed. Such behaviours drive product life-cycles, intensify competition to establish technical standards that dominate a market \citep{shapiro1998information}, and underpin the innovator's dilemma that favours incrementalism within a established trajectory against the exploration of radical alternatives \citep{christensen2013innovator, march1991exploration}. The search for efficiency also strengthens processes of institutional isomorphism' leading organisations in a sector to adopt similar processes, routines and organisational structure to facilitate coordination of productive activities \citep{dimaggio1983iron, nelson2009evolutionary}.

% Collaborating with industry is one of the main channels through which academic researchers are able to access the data and infrastructure required for state-of-the-art deep learning research \citep{wahmed2020}. 
It is hard to deny that large technology companies are playing a pivotal role in contemporary AI research: they have produced some of its biggest breakthroughs, recruit large  numbers of researchers and graduates from leading academic institutions, collaborate with academic researchers who in many cases have dual academic - industrial affiliations \citep{gofman_brain_drain, hain2020,hagendorff2020ethical, lundvall_chinas_2022} and sponsor popular open source packages for the implementation of deep learning methods. They also participate actively in key conferences such as NeurIPS (Neural Information Processing Systems) or ICML (International Machine Learning Conference).\footnote{\href{https://medium.com/@dcharrezt/neurips-2019-stats-c91346d31c8f}{https://medium.com/@dcharrezt/neurips-2019-stats-c91346d31c8f}} There are multiple reasons explaining why private companies have adopted this open innovation and publishing strategy despite the risk of spillovers that could benefit their competitors \citep{arora2021knowledge}. They include complementarities between AI techniques and their proprietary data / infrastructures \citep{bessen2017information, rikap_tech_2021}, the desire to attract talent \citep{hain2020} and lack of competition between technology giants operating in adjacent markets.

% One exception to the predominance of private firms at the cutting-edge of contemporary AI research is OpenAI, a not-for-profit lab that has the the mission of `ensuring that AI benefits all of humanity'.\footnote{\href{https://openai.com/about/}{https://openai.com/about/}}- but even OpenAI's modus operandi has over, time, converged with those of private sector labs, receiving large investments from Microsoft and commercialising the outputs of GPT-3, a powerful text generator it has developed \citep{haomessy2020}.

Is there any evidence that this is narrowing AI research? 

% These companies have been behind key recent breakthroughs in deep-learning driven AI research, sponsor the development of open source frameworks such as TensorFlow and Pytorch to encourage the uptake of deep learning techniques across the AI ecosystem, release open datasets and offer access to cloud infrastructures to scale up AI analyses - this means that they strengthen the dominant deep learning trajectory directly through their own research, and indirectly, by providing complementary data, software and computational resources that others can use to deploy - but not necessarily advance the state-of-the-art of - these methods.

\cite{wahmed2020}'s analysis of computer science research suggest that private sector participation has created a `compute divide' between them and their elite collaborators on the one hand, and smaller institutions on the other, potentially leading to what they refer as a `de-democratisation' of AI research. \cite{whittaker2021steep} claims that corporate investments in collaborative research with universities have `captured' academic researchers and skewed their research agendas. \cite{bender2021dangers} analysis of the ethical risks raised by large language models developed by companies such as Google states that industry's focus on those models creates an opportunity cost  \textit{"as researchers pour effort away from directions requiring less resources"} - in other words, in directions distinct from the dominant deep learning AI design. Google's opposition to the publication of this paper, which eventually resulted in the dismissal of Timnit Gebru and Margaret Mitchell, two of its leading ethics researchers, suggests that private companies have strong incentives to downplay the limitations and weaknesses of AI systems based on deep learning, thus reducing the incentives to explore alternatives.

\section{Research questions, data and methods}\label{sec:data}

\subsection{Research questions}
Our literature review has provided a theoretical and qualitative basis for a potential decline in AI's thematic diversity which creates the risk of lock-in to a dominant deep learning design with important limitations. It also points at the potential role that large technology companies may be playing in this process, directly through their R\&D activities, and indirectly through their influence in the AI research ecosystem. The rest of the paper sets out to provide empirical evidence about these processes through a descriptive analysis to address the following research questions:

\begin{enumerate}
    \item Is AI research becoming thematically narrower?
    \item How does the thematic diversity of private sector organisations compare with those in academia and the public sector?
    \item How influential is private sector AI research?
\end{enumerate}

\subsection{Data sources}
In order to address these research questions, we create a novel dataset that combines information from arXiv, Microsoft Academic Graph (MAG) and the Global Research Identifier. 

% Table \ref{tab_2_vars} presents key variables.

% \begin{table}
%     \centering
%     \small
%     \begin{tabular}{p{0.9in} p{0.5in} p{0.8in} p{3in}}
%     \toprule
%     Table & Source & Variable & Definition \\
%     \toprule
%     \texttt{article} & arXiv & title & Article title \\ \\
%     \texttt{article} & arXiv & Created & Date when the article was created \\ \\
%     \texttt{article} & arXiv & categories & arXiv categories (scientific and technical sub-disciplines) that the articles have been labelled with (Can be more than one per article) \\ \\
%     \texttt{article} & arXiv & abstract & Article abstract \\ \\
%     \texttt{institution} & MAG & institution & Institutional affiliation of article authors (set) \\ \\
%     \texttt{institution} & GRID & Institution type & Can be Company, Education, Facility, Government, Nonprofit, Healthcare, Other \\
%     \bottomrule
%     \end{tabular}
%     \caption{Summary of key variables in the data}
%     \label{tab_2_vars}
% \end{table}

\subsubsection{arXiv}
arXiv is an online pre-print repository that in recent years has become an important outlet for the dissemination of AI research results close to real time.\footnote{\url{https://www.arxiv.org}}.  We have downloaded the whole arXiv corpus, comprising 1.84 million articles published between April of 1986 and August 2020 through arXiv's API.\footnote{The data has also recently become available as a bulk download from machine learning competition site Kaggle (\url{https://www.kaggle.com/Cornell-University/arxiv}).} For each article, we obtain its id, title, date when it was created, its categories (the arXiv categories with which the authors label a article with when they submit it reflecting the scientific discipline or sub-discipline that it belongs to), its abstract and its citation count.

\subsubsection{Microsoft Academic Graph (MAG)}
We obtain the institutional affiliation of an article's authors from Microsoft Academic Graph (MAG), a large scientometric database collected and enriched by Microsoft Cognitive Services which is particularly suitable for the analysis of computer science and engineering disciplines \citep{wang2020microsoft, martin2021google}. We query MAG with the title of the arXiv articles (see Klinger et al (2018) for additional details about this approach) and extract the institutional affiliation of each article author as of the time of publication.\footnote{MAG uses text mining to automatically extract author affiliations from a paper's metadata, which includes text processing tasks such as entity disambiguation \citep{wang2020microsoft}.}

\subsubsection{Global Research Identifier (GRID)}
We fuzzy match the institute names extracted from Microsoft Academic Graph with the Global Research Identifier (GRID), a public database with detailed metadata about research institutions globally in order to identify their type (Company, Education etc).\footnote{\url{https://www.grid.ac/}} We do this using the same algorithm as \citet{klinger2018deep}, which combines multiple fuzzy matching methods to identify GRID institutes that have the same names as institutions in MAG. This yields just over one million articles with institutional information and 2.45 million article-institute pairs. There is a large number of articles with missing institutions in the early years of arXiv operation (where as we will show AI activity was very limited) and slight growth of articles without matched institutes in recent years. Visual checks of the articles lacking institutional information suggests that this is because this information is not available from MAG or because the research involves less well-known institutes in some cases based in developing countries.

One issue with our approach is that it generates duplicate matches for multinational institutions with presence in multiple locations (for example, \texttt{Google}, a single institute in MAG, is matched with multiple GRID institutes including \texttt{Google (United States)}, \texttt{Google (United Kingdom)} etc.). To address this we split name strings on parentheses, retain the organisation name and remove duplicate article-organisations observations. This means that in the analysis that follows we do not consider how many times a single institution participates in an article, but simply that it does.

After inspecting the data manually we identify mis-classified / missing data about two notable participants in AI research: DeepMind's papers are classified in `Google' by MAG, and OpenAI is not included in GRID. To address this, we scrape the research section of the websites of both organisations, extract the arXiv ids of their papers and reclassify the papers in our dataset. In the rest of our analysis we assume that DeepMind papers do not involve Google researchers although we know this is not always the case.

\subsubsection{Observations about potential biases}
arXiv pre-prints are not subject to peer review, potentially raising concerns about their quality. It is however worth noting that there are some minimum quality thresholds to publish in the platform: submissions are reviewed for relevance and new authors are validated by existing participants. Previous work has also shown that the majority of papers submitted to the prestigious AI conference NeurIPS are posted in arXiv, and found a strong correlation between the geography of deep learning research in arXiv and peer-reviewed AI publications as well as the location of data-oriented technology startups \citep{klinger2018deep}. This suggests that variation in AI research in arXiv is associated with AI research in other publication channels as well as with other technological and entrepreneurial activities related to AI. As our findings will show, leading businesses and academic institutions across the world use arXiv to share their AI research suggesting that it is relevant for our analysis. 

One potential advantage of using arXiv over peer-reviewed publications is that it may contain more publications by companies who want to disseminate findings rapidly ahead of conferences or product launches, but are less driven by the desire / incentives to achieve scholarly esteem via peer-reviewed publications, which is of interest to us. The fact that arXiv data is not subject to peer review filters could also mean that it reflects better the diversity of views and techniques present in AI research, admittedly including some off-kilter methods and eccentric studies. At the same time, our data may exclude research results that private sector organisations choose not to disclose publicly \citep{arora2021knowledge}. We would expect this to exclude more downstream / applied findings which would have reduced the thematic diversity of our corpus.

Our matching strategy may exclude smaller and less well-known research institutions, with ambiguous implications for our analysis of diversity: while one would expect startups to develop more radical ideas that could increase the thematic diversity of the private corpus of AI research, this is likely to involve low volumes of activity and influence. We would also expect startups to have fewer resources to invest in basic research that sets the agenda of the field, one of our main interests. In parallel, the exclusion of research institutions outside of the AI mainstream could also bias our estimates of the thematic diversity of AI research in academia downwards. Our analysis of the evolution of diversity in AI research considers the full corpus, so it will not be biased by any coverage gaps created by our matching process.

\subsection{Methods}

\subsubsection{Procedure to identify AI articles}
\label{subsubsec:identify}

Recent AI mapping studies have used keyword searches and topic modeling of abstracts to identify relevant terms \citep{cockburn2018impact, klinger2018deep, stathoulopoulos2019gender, frank2019evolution, bianchini2020deep}. Here, we are interested in identifying articles from previous eras of the history of AI that we are less familiar with, and which may have a weaker presence in the corpus, making them harder to detect in a single arXiv category or using topic models trained on all the corpus.

In order to address this, we exploit the fact that arXiv preprints are labelled by disciplinary category, and some of these are clearly related to our field of interest. We refer to them as the `core AI categories'. They are \textit{cs.AI} (Computer Science: Artificial Intelligence), \textit{cs.NE} (Computer Science: Neural and Evolutionary Computing), \textit{cs.LG} (Computer science: Machine Learning) and \textit{stat.ML} (Statistics: Machine Learning)). In total, they comprise just under 89,000 unique articles  (there are 139,000 articles with at least one of the categories but there are significant overlaps between them).  One risk of focusing exclusively on these categories is that this could lead us to exclude important applications of AI in other fields of STEM and computing. The challenge is how to incorporate systematically into our corpus such AI applications core AI categories.

% Previous research about deep learning has for example shown that arXiv categories such as \textit{cs.CV} (Computer Vision) and \textit{cs.CL} (Computer Language) have been important sites for AI development and deployment \citep{klinger2018deep}. 

To do this, we consider core AI categories as research areas that specialise in the development of AI techniques that are then applied by researchers in other fields. We would expect terms related to those AI techniques to be over-represented (salient) in core AI categories, and present in other articles that adopt them. We have developed a procedure that identifies and expands those terms in core AI categories and then identifies other articles outside those categories with unusually high occurrences of those terms (see the technical annex for additional information about this approach).

Since several parameters in our AI identification procedure have been selected heuristically, we explore an alternative strategy to identify AI papers that expands our semantic approach with citation information (see technical annex for a description of this approach) and analyse its impact on our results in Subsection \ref{subsubsec:robust_evolution}.

\subsubsection{Topic modelling}
We use topSBM, a hierarchical topic modelling algorithm, to estimate the thematic composition of AI research, which we will then use to calculate various metrics of diversity. TobSBM adopts a network approach to estimate topics in the data \citep{gerlach2018network}. This involves transforming the pre-processed corpus of $n$ article abstracts into a network where different words are connected through their co-occurrence in abstracts. This network is decomposed into communities using the stochastic block model (SBM), a generative Bayesian model for random graphs \citep{abbe2015exact}. 

This results in a collection of $k$ topics $T = \{t_1,...,t_k\}$ where each topic $i$ has a word mix $W_i = \{w_{i,1}, w_{i,2}, ...,w_{i,s}\}$ representing the probability that a word belongs to it and ach article $l$ has a topic mix $P_l = \{p_{l,1},...p_{l,k}\}$ representing the probability that a topic is present in the article.

In the rest of the paper we will consider those topic distributions over a paper as proxies for the themes that constitute it and, when aggregated, as measures of the thematic composition of particular sub-corpora of AI research, such as all AI research in a year, or all AI research in private sector companies or particular organisations. 

TopSBM has some advantages over other topic modelling methods and in particular the widely adopted Latent Dirichlet Allocation algorithm \citep{blei2003latent}. First, it makes weaker assumptions about the data-generating processes: LDA assumes that topics in a corpus are generated by the Dirichlet distribution and tends to under-perform when this assumption is violated. TopSBM relies on more general priors making it more robust to heterogeneous topic distributions. Second, tobSBM uses Bayesian inference to automatically identify the number of topics that maximises the posterior distribution of links between communities (which define topics) and yields the shortest description of the data while reducing the risk of overfitting \citep{peixoto_2017}. This removes the need to tune the number of topics manually using hard to interpret tests and heuristics, a common problem with LDA. TopSBM has been shown to outperform LDA in tests with simulated and real-world corpora.

Notwithstanding these advantages, it is still risky to base all our analysis on a single algorithm and automated procedure to estimate a key variable. We therefore assess the robustness of our findings to adopting an alternative topic modelling strategy (LDA) with a variable number of topics in subsection \ref{subsubsec:robust_evolution}.

When we fit the topic model on an AI corpus pre-processed using the same approach we used for our AI identification procedure (and described in the technical annex), it yields 750 topics from which we remove 193 topics that appear in more than 10\% of the corpus, and generally contain generic and uninformative collections of terms frequently used in academic research such as \textit{`using demonstrates previously reported'}, \textit{`recent previous variety ones demonstrating'} or \textit{`found initial investigated analyzed investigation'}, and hard to interpret topics comprising two words or less. We present examples of the topics underpinning our analysis in various tables across the paper and make the full list of topics available in the project's GitHub repository.%add-link

% \footnote{The topic model is fitted on the same pre-processed - tokenised and tri-grammed - corpus that we described in \ref{subsubsec:identify}.} 

\subsubsection{Metrics of diversity}
In general, diversity refers to the heterogeneity of the elements of a population in relation to some class that takes different values, such as for example the species in an ecosystem, the industries in an economy, or the ethnicity of a population \citep{page2010diversity}. This heterogeneity can be measured along the dimensions of \textit{variety} (capturing the number of classes that are present in the population), \textit{balance} (capturing whether the proportions of different classes are evenly distributed) and \textit{disparity} (capturing the degree of heterogeneity between the classes in the population). A population with more classes, a more even distribution of sub-populations between classes, and with classes that are more heterogeneous will be more diverse. 

In Section \ref{subsec:div_evol} we take the corpus of AI research in arXiv as our population, and research ideas and techniques belonging to different topics as the classes. We use these to calculate three metrics of diversity that put different emphasis on the dimensions above. They are a metric of balance that considers the concentration of AI research on different topics, Weitzman's metric of ecosystem diversity based on overall distance (disparity) between topics in the corpus \citep{weitzman1992diversity} and the Rao-Stirling metric of diversity, that takes into account the balance and disparity of topics in the corpus \citep{stirling2007general}.

By using these three metrics in parallel, and parametrising them in different ways, we aim to ensure that our findings are robust to various definitions and operationalisations of diversity, and also to increase their interpretability by considering them from different perspectives. Tables \ref{tab5_div_def} and \ref{tab6:param} respectively define our metrics and the parametres we have used to operationalise them.

\begin{table}[]
    \centering
    \small
    \begin{tabular}{p{0.75in} p{2.2in} p{2.2in}}
    \toprule
    Metric & Definition & Operationalisation  \\
    \toprule
    Balance & Distribution of classes & Hirschmann-Herfindahl (HH) index and Shannon entropy based on the share of all topics or articles accounted by a topic \\ \\
    Weitzman & Sum of distances of the dendrogram representing a hierarchical clustering algorithm trained on the data & Distance measures based on topic co-occurrence in articles \\ \\
    Rao-Stirling & Product of shares of classes by their pairwise distances & Shares based on topic presence in corpus or article, distances based on topic co-occurrence in articles \\ \\
    \bottomrule
    \end{tabular}
    \caption{Definition and operationalisation of metrics of diversity}
    \label{tab5_div_def}
\end{table}

\begin{table}[]
    \centering
    \small
    \begin{tabular}{p{0.6in} p{1.5in} p{1.5in} p{1.5in}}
    \toprule
    Metric & Parametre Set 1 & Parametre Set 2 & Parametre Set 3  \\
    \toprule
    Balance & HH index of topic distribution over population of topics present in the corpus & HH index of topic distribution over articles assigned to their top topic &
    Shannon entropy of topic distribution over population of topics present in the corpus \\ \\
    Weitzman & Cosine distance between topics & Chebyshev distance between topics & Jaccard distance between topics (binarising on topic presence) \\ \\
    Rao-Stirling index & Rao-Stirling index of topic distribution over population of topics present in the corpus with a threshold above 0.1 and cosine topic distance & Rao-Stirling index of topic distribution over population of topics present in the corpus with a threshold above 0.1 and correlation topic distance & Rao-Stirling index of topic distribution over articles assigned to their top topic and cosine topic distance
    \\ \\
    \bottomrule
    \end{tabular}
    \caption{Parametre set details by diversity metric}
    \label{tab6:param}
\end{table}

There are some important differences between our metrics that will bear on their interpretation. The balance and Rao-Stirling metric are similar in that they both take into account the distribution of topics in the population, in the case of Rao-Stirling weighted by the distance between topics that we measure using pairwise distances in the topic distance matrix, with topics closer within that distance matrix (based on their co-occurrence in articles) appearing closer to each other. This means that if there are many topics in the corpus but most of them account for a small share of the total, this will reduce the corpus' diversity. Weitzman diversity, by contrast, does not take into account the distribution of topics in the corpus but only their distance (again based on co-occurrences). This means that it would be possible to have an extremely concentrated topic distribution with very high diversity as long as there is a long tail of minority topics that are very different from the dominant ones. In a way one could think of the Weitzman metric of diversity as an indicator of the diversity of classes hosted in a population rather than its actual manifestation in terms of the importance of these classes. In his economic analysis of ecological diversity, Weitzman complemented an initial analysis of diversity using this metric with an analysis of the economic cost of preserving it that took into account the size of different classes - we do not take that second step here \citep{weitzman1993preserve}.

% \subsubsection{Sentence embeddings}

% In the last part of our analysis we will consider the position of organisations participating in AI research in a semantic space that we map using BERT (Bidirectional Encoder Representations and Transformers), a deep learning technique that learns vector representations of text based on their context \citep{Devlin2019}. We train this model on the 1.8 million abstracts in the arXiv corpus, which yields a 674-dimensional representation of each article. We calculate the average of these vectors for the articles involving organisations of interest and visualise their positions using tSNE (t-Stochastic Neighbour Embedding), a dimensionality reduction technique that projects high-dimensional data in a 2-d or 3-d space \citep{maaten2008visualizing}.

\section{Findings} \label{sec:result}

\subsection{Evolution of AI research and its thematic diversity}
\label{subsec:ai_evol}

\subsubsection{AI research trends}

We begin by considering the evolution of AI research in arXiv. Is the widespread perception of an `AI boom' reflected in our data? 

\begin{figure}[!htbp]
    \centering
    \includegraphics[width=0.9\textwidth]{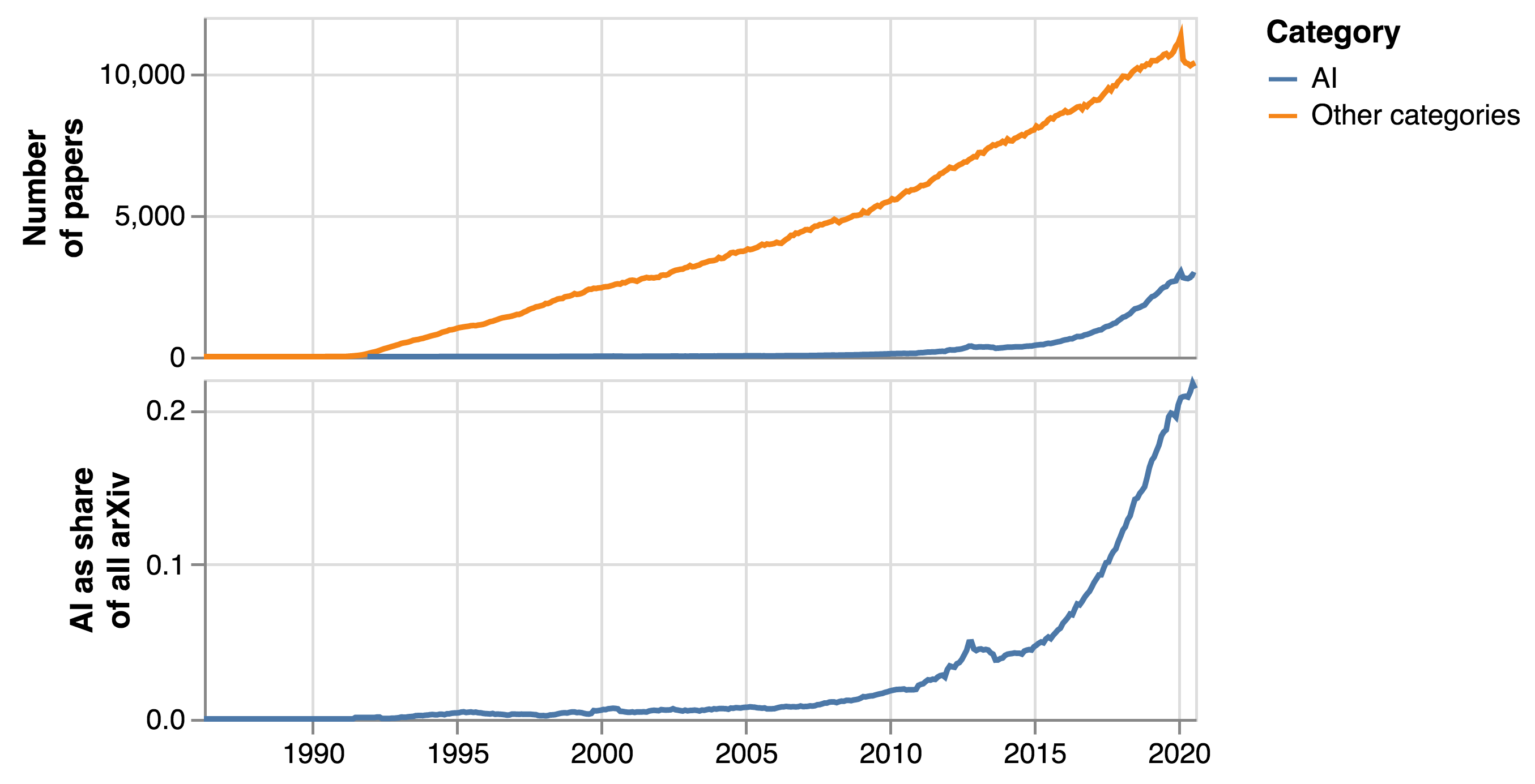}
    \captionsetup{margin=1cm}
    \caption{The top panel shows the total number of monthly articles posted in arXiv in AI and all other categories. The bottom panel shows AI articles as a share of the arXiv corpus. Both series are smoothed using rolling averages.}
    \label{fig:agg_trends}
\end{figure}

Figure \ref{fig:agg_trends} shows the monthly number of papers published in AI research and the rest of the arXiv corpus (top), and the share of the arXiv corpus accounted by AI papers (bottom). By contrast to non-AI categories, where the number of articles has grown at a steady pace since the 1990s, growth in AI research has been more recent, and taken place very rapidly. Remarkably, in the first half of 2020 AI articles comprised 20\% of all publications uploaded to arXiv. 94\% of all the AI papers in our corpus were published after 2012, and 60\% after 2018. 

Has rapid growth of activity in arXiv has been accompanied by shifts in its topical composition? To visualise the answer to this question, we have assigned each topic in our data to the arXiv category where it has the highest salience.\footnote{It is important to point out that we are assigning topics to arXiv categories with the goal of interpreting the evolution of AI research and the areas of specialisation of different organisations in section \ref{subsec:priv_spec} but it does not play a role in our analysis of diversity, which is based on the raw outputs of our topic model.}

We calculate this analogously to a location or revealed comparative advantage quotient $Q_{i,c} = \frac{s_{i,c}}{s_{i}}$ where $s_{i,c}$ is the share of topic $i$ in the category $c$ and $s_i$ is the share of the topic in the corpus. $Q_i > 1$ indicates that a topic is over-represented in a category.  Table \ref{tab:topic_examples} includes some examples of salient topics randomly drawn from various arXiv categories, illustrating their thematic focus: AI papers in \textit{stat.ML} refer to statistical techniques such as Gaussian processes and Bayesian inference while \textit{cs.CR} (Cryptography and Security) contains topics about software vulnerabilities, privacy, adversarial attacks against deep learning systems and authentication. \textit{cs.CY} addresses a range of societally oriented topics related to health (and more specifically pandemics, including AI research on Covid-19), education and ethical issues.

\begin{figure}[h!]
    \centering
    \includegraphics[width=0.9\textwidth]{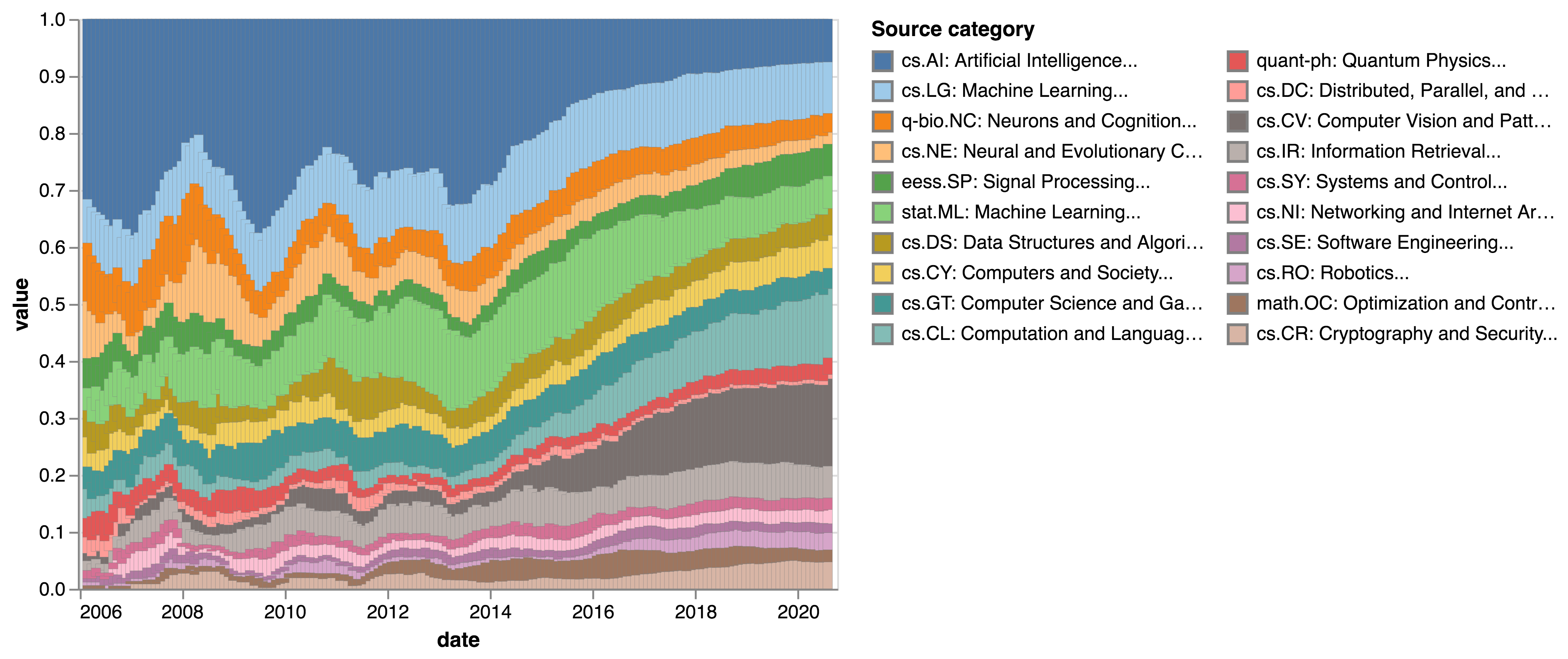}
    \captionsetup{margin=1cm}
    \caption{Evolution of topical composition of AI research: each topic has been labelled with the arXiv category where it has the strongest representation. The population each year is the sum over categories of all topics occurrences over 0.1 in a paper.}
    \label{fig:topic_trends}
\end{figure}

In Figure \ref{fig:topic_trends} we visualise the evolution of the relative importance of topics associated to different arXiv categories as the share of articles presenting a topic in each category over all topic occurrences in all articles.\footnote{We assume that a topic is present in an article if its weight on it is above 0.1}

Consistent with the narrative of a `deep learning' boom, it shows rapid growth in the importance of topics salient in \textit{cs.CV} (Computer vision) and \textit{cs.CL} (Computer Language), two domains with an abundance of unstructured data where deep learning has contributed to important advances. Since the mid 2010s, \textit{cs.CR} (Cryptography and security) has also gained prominence - this is linked to growing interest in the risk of adversarial attacks, as well as privacy and cybersecurity. Other topical areas that have seen growth include \textit{cs.CY} (Computer and Society), with research about AI applications in health, its educational aspects and ethical implications, \textit{cs.IR} (information retrieval), linked to the development of search engines and recommendation system, an important AI application area for technology companies, and \textit{cs.RO} (robotics).  Meanwhile, we see a decline in the relative importance of topics related to \textit{cs.AI}, which tend to be more focused on symbolic techniques and slight decline (specially since the mid 2010s) of \textit{stat.ML} (statistical machine learning) topics involving various machine learning techniques outside of deep learning.

Interestingly \textit{cs.NE} has seen a decline since the 2010s despite the increasing popularity of `neural computing' techniques. One explanation for this result is that many deep learning-related techniques have been developed in application domains such as computer vision and computer language.

% Figure \ref{fig:topic_trends} also suggests qualitatively that the thematic diversity of AI research has increased over time, with less concentration of topical activity in a small number of categories. The figure tells us little, however, about the thematic composition of AI at the level of individual topics, or about the extent to which the topics being pursued have high disparity, that is, are significantly different from each other. We will come back to these questions in Section \ref{subsec:div_evol}. 

\begin{table}[]
    \centering
    \small
    \begin{tabular}{p{0.5in} p{5in}}
    \toprule
    Category & Examples \\
    \midrule
cs.AI & artificial\_intelligence\_artificial\_computer\_intelligence\_life... \newline machine\_machines\_pattern\_recognition\_steady\_turing\_machine... \newline describe\_elements\_description\_forms\_element... \newline proof\_proofs\_theorems\_automated\_reasoning\_conjectures... 
\\ \\
cs.NE & capacity\_coefficients\_coefficient\_analytically\_permutation... \newline development\_engineering\_prototype\_creation\_organization... \newline describing\_chapter\_ais\_dca\_som... \newline building\_paper\_presents\_modelling\_built\_integration... 
\\ \\
cs.LG & emph\_algorithmic\_line\_satisfy\_arising... \newline arms\_arm\_thompson\_sampling\_bandit\_bandits... \newline loss\_losses\_cross\_entropy\_cross\_entropy\_loss\_sharpness... \newline memory\_stored\_store\_storing\_memories... 
\\ \\
stat.ML & gaussian\_process\_gaussian\_processes\_gps\_inducing\_gaussian\_process\_regres... \newline independent\_mean\_dependent\_fraction\_moment... \newline signal\_signals\_signal\_processing\_snr\_complex\_valued... \newline bayesian\_likelihood\_posterior\_bayesian\_inference\_posterior\_distribution... 
\\ \\
cs.CV & style\_art\_sketch\_styles\_sketches... \newline source\_domain\_adaptation\_target\_domain\_cross\_domain\_unsupervised\_domain\... \newline tracking\_tracker\_speaker\_verification\_front\_end\_speaker\_recognition... \newline spectral\_spectra\_laplacian\_band\_bands... 
\\ \\
cs.CL & rare\_drug\_medicine\_biomedical\_drugs... \newline natural\_language\_meaning\_symbolic\_descriptions\_compositional... \newline sequence\_sequences\_hmm\_hidden\_markov\_variable\_length... \newline vector\_vectors\_proximity\_inner\_product\_dot\_product... 
\\ \\
cs.CR & security\_iot\_secure\_protection\_vulnerabilities... \newline privacy\_private\_differential\_privacy\_differentially\_private\_privacy\_prese... \newline adversarial\_adversarial\_examples\_adversarial\_training\_adversarial\_attacks\... \newline authentication\_fingerprints\_biometric\_fingerprint\_pain... 
\\ \\
cs.IR & feature\_selection\_classification\_regression\_elm\_linear\_discriminant\_analy... \newline items\_recommendation\_recommendations\_item\_recommender\_systems... \newline emotion\_emotions\_emotional\_stress\_emotion\_recognition... \newline retrieval\_hashing\_image\_retrieval\_triplet\_triplet\_loss... 
\\ \\
cs.CY & infection\_pandemic\_epidemic\_infected\_virus... \newline program\_programs\_programming\_induction\_syntax... \newline students\_course\_education\_educational\_university... \newline law\_society\_legal\_ethical\_stakeholders... 
\\
    \bottomrule
    \end{tabular}
    \caption{Random salient topics in selected arXiv categories}
    \label{tab:topic_examples}
\end{table}

\subsubsection{Evolution of thematic diversity}
\label{subsec:div_evol}
Has the change in the thematic composition of AI that we just described been associated to an increase or decrease in its thematic diversity?

\begin{figure}[h!]
    \centering
    \includegraphics[width=0.9\textwidth]{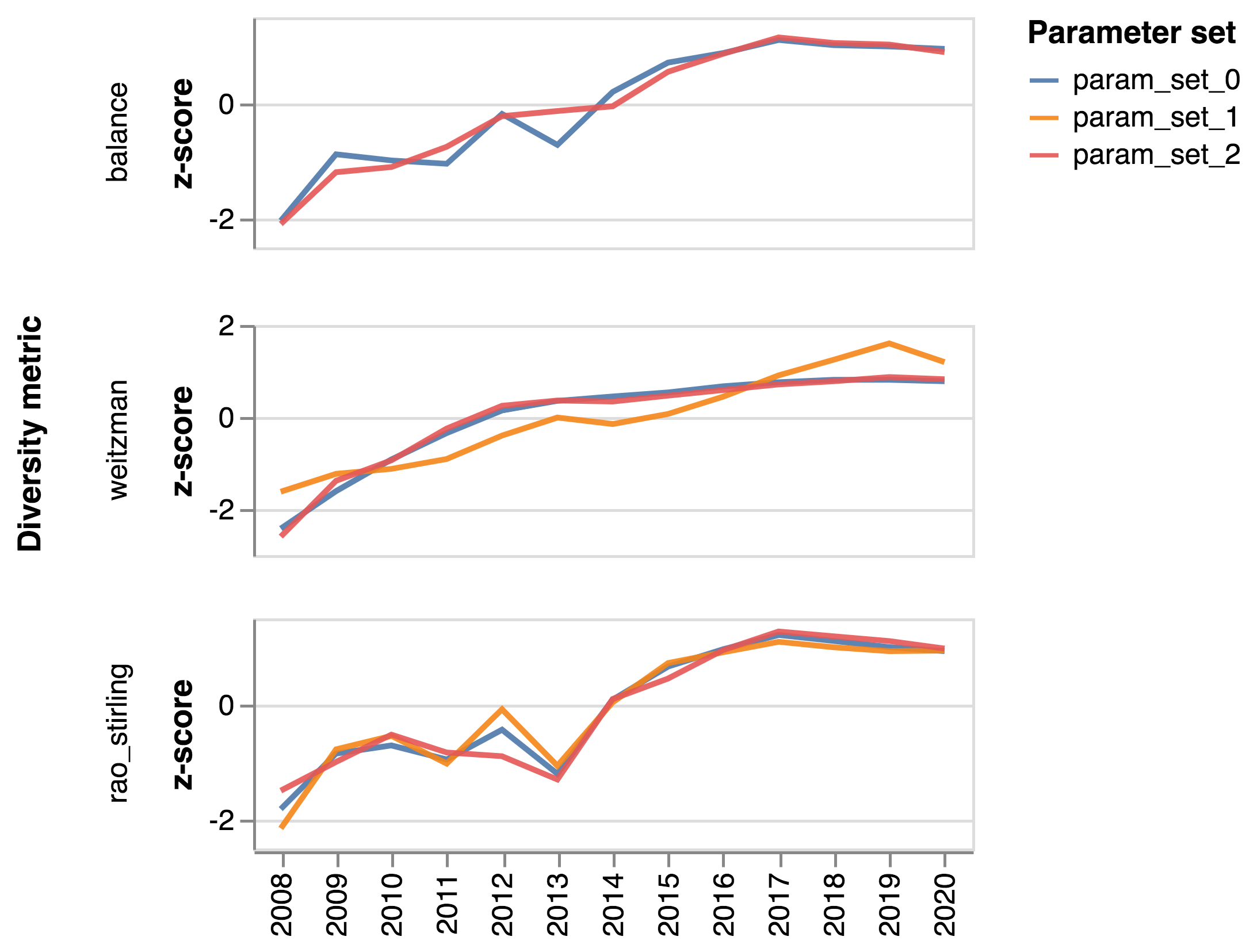}
    \captionsetup{margin=1cm}
    \caption{Evolution of thematic diversity in AI research. Each panel presents the evolution of diversity according to one of our metrics (balance, Weitzman and Rao-Stirling) and different parametre set. We have calculated the z-score inside each of our time series to make them comparables across parametre sets yielding different absolute values.}
    \label{fig:div_evol}
\end{figure}

Figure \ref{fig:div_evol} presents the evolution of this variable based on our three diversity metrics and three parametre sets. 
% In rough terms, all our operationalisations of diversity present a similar picture, with an initial increase in diversity followed by stabilisation / stagnation and even a slight decline for some of the metrics/parametre sets in recent years.

The balance metric (which focuses on the distribution of activity over topics) and the Rao-Stirling metric (which weights balance by the distance between topics) present very similar trends, with a strong increase of diversity after 2012 followed by stagnation from 2017 onwards. The Weitzman metric, which considers the total distance between topics in the corpus but not their concentration is smoother, and its temporal sequence is different, with earlier growth and stagnation of diversity or (in the case of parametre set 1 where we calculate topic distance using the Chebyshev metric) a recent increase of diversity followed by a drop in 2020.\footnote{This decline may be linked to the fact that our 2020 corpus does not include a full year of activity, reducing the number of topics present in the topic and therefore the sum of distances between them which is used to calculate the Weitzman metric)}.

The technical annex presents additional analyses of the evolution of distance between AI research topics in the topic co-occurrence network.

\subsubsection{Robustness tests}
\label{subsubsec:robust_evolution}

In section \ref{sec:data} we identified several sources of potential bias in our analysis: the procedure we have used to identify AI papers includes several parameters chosen heuristically and we rely on a single topic modelling algorithm to estimate the thematic composition of the AI corpus. An additional concern is that the strong concentration of AI research in recent years may skew the results of our topic model by favouring research themes and topics that have appeared more recently, thus distorting the results of our longitudinal analysis.

Here, we implement three robustness tests to asses if these factors are impacting our results.

\begin{enumerate}
    \item We adopt a new definition of AI that includes papers that were cited from / to a sample of 10,000 papers in our corpus (see technical annex for additional details). We focus on those papers that were cited to / from the corpus more than once combined with a sample of AI papers from our corpus.\footnote{ (we do not include the full sample for reasons of computational efficiency and to avoid creating an imbalanced corpus dominated by the papers we identified originally).} We then fit a topic model on this corpus in twenty replications with a topic number randomly drawn between 200 and 400, and estimate all metrics of diversity and parameter sets.
    \item We reproduce our analysis on a sample of 50,000 papers from our original AI corpus replacing topSBM with an LDA model in twenty replications with the topic number randomly drawn between 200 and 400 topics, and estimate all metrics of diversity and parametre sets on these topic distributions.
    \item We build a temporally balanced corpus where we draw 1000 papers from our original corpus each year between 2010 and 2020. We then train an LDA model on this corpus with a topic number randomly drawn between 200 and 400 in twenty replications, and estimate all metrics of diversity and parametre sets in the resulting topic distributions.
\end{enumerate}

\begin{figure}[h!]
    \centering
    \includegraphics[width=0.9\textwidth]{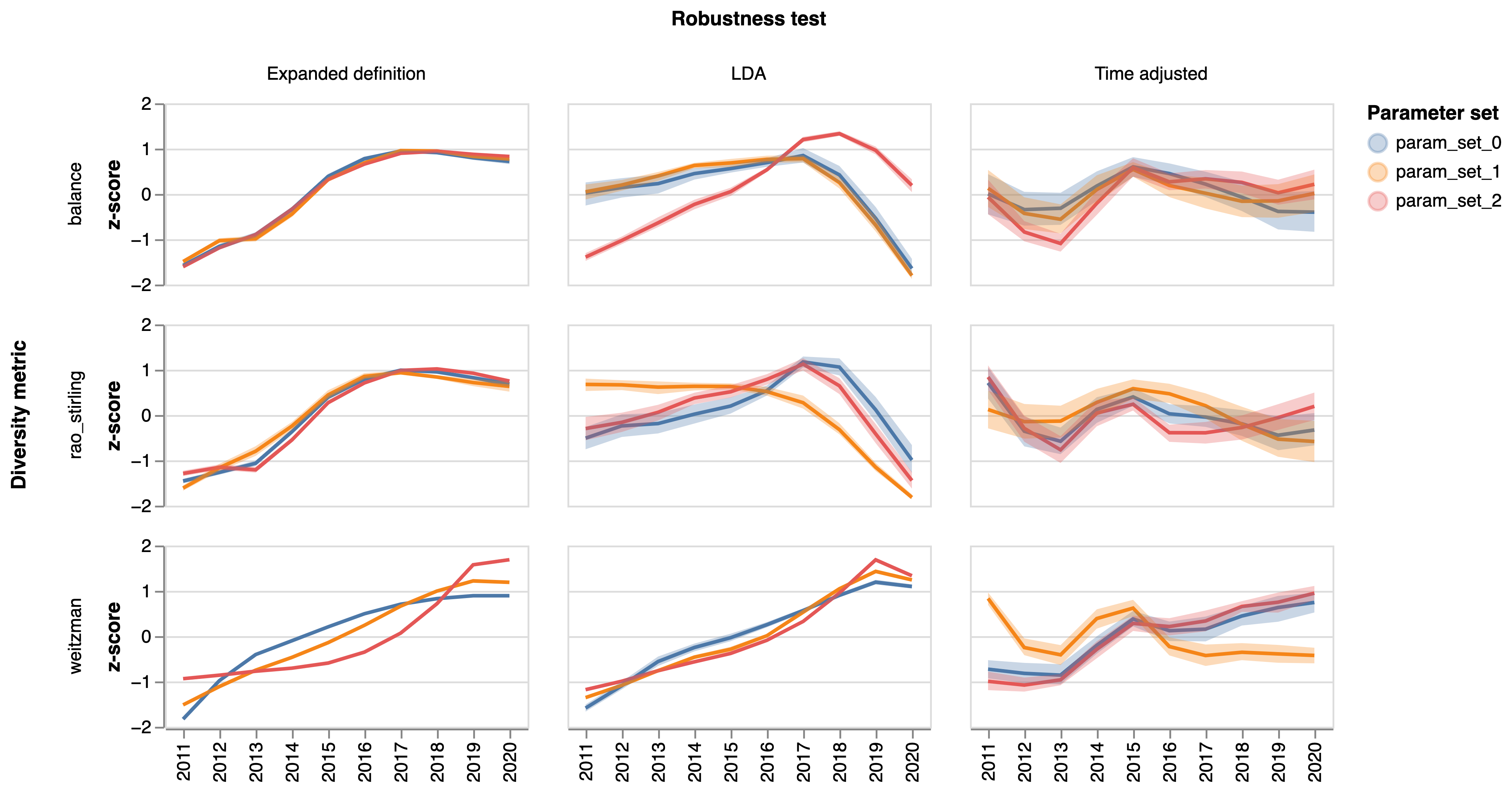}
    \captionsetup{margin=1cm}
    \caption{Each column represents a robustness test (use of an expanded definition of AI, use of an LDA topic model with variable number of topics, use of a corpus that samples the same number of papers per year) and each row represents a diversity metric. The colours capture the parametre set used. We report average scores and confidence intervals for twenty replications per robustness test, diversity metric and parameter set.}
    \label{fig:rob_test}
\end{figure}

Figure \ref{fig:rob_test} presents the results for each of our replication tests, diversity metrics and parametre sets. We find that the results are generally consistent with the analysis presented before: an initial increase of diversity in the mid 2010s followed by stagnation and in some cases decline. As before, the patterns are different in some of the Weitzman metrics (particularly when estimated using the LDA model), which we link to the fact this metric does not take into account topic concentration. The results we reported previously do not appear to be significantly altered by a change in the method we use to identify AI papers or by the temporal distribution of AI activity in our corpus, and still suggest a tendency towards a narrowing in the thematic diversity of AI research in recent years. 

\subsection{Thematic diversity in the private sector}
\label{subsec:ai_evol}

Having studied the evolution of thematic diversity in AI research, we move on to consider differences between the thematic diversity of organisations in the private and public sector, thus addressing our second research question. We begin this by considering the levels of participation of private companies in AI research - to this extent is the high level of participation and influence discussed in section \ref{subsec:lit_private} visible in our data?

\subsubsection{Private participation in AI research}

\begin{figure}[h!]
    \centering
    \includegraphics[width=0.9\textwidth]{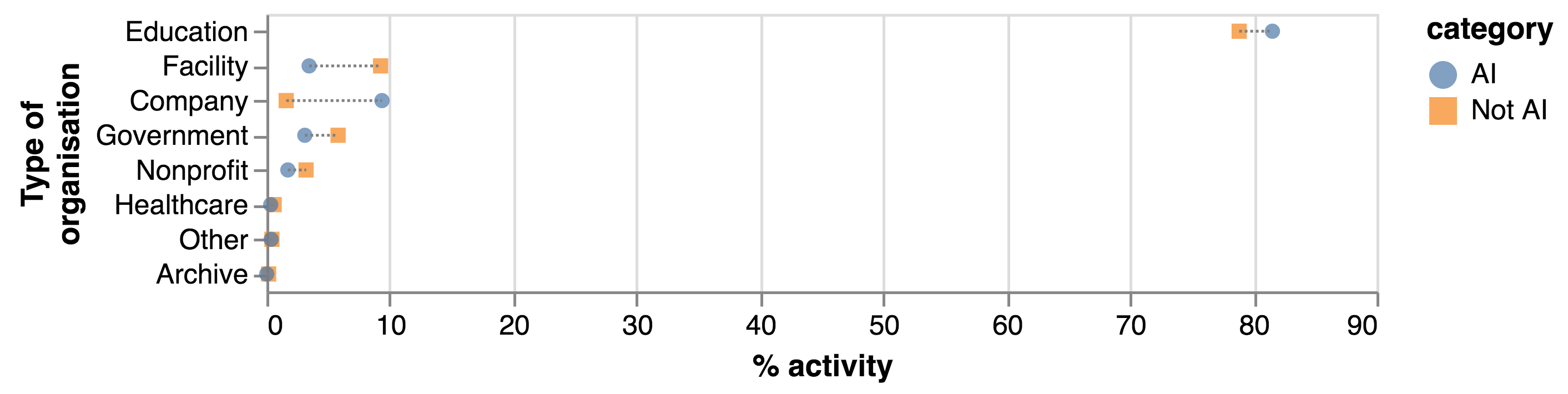}
    \captionsetup{margin=1cm}
    \caption{Share of participation of different organisation types in AI research compared to the wider arXiv corpus. Participation is measured as the share of organisation type participation (number of articles) normalised by all research participations}
    \label{fig:org_part}
\end{figure}

Figure \ref{fig:org_part} compares the level of participation of various types of organisations in AI research with the situation in other research areas in arXiv, calculated as the total number of articles involving an organisation type normalised by the number of articles involving all organisations. It shows, perhaps unsurprisingly, that academic institutions (\textit{Education}) are most active in the corpus. We also find support for the idea that the private sector is particularly active in AI research: private companies account for around 10\% of all research participations, a share ten times higher than their participation in research outside of AI. By contrast, Government, nonprofit and Facilities are less active in AI research compared to their shares of participation in the wider corpus.

\begin{figure}[h!]
    \centering
    \includegraphics[width=0.9\textwidth]{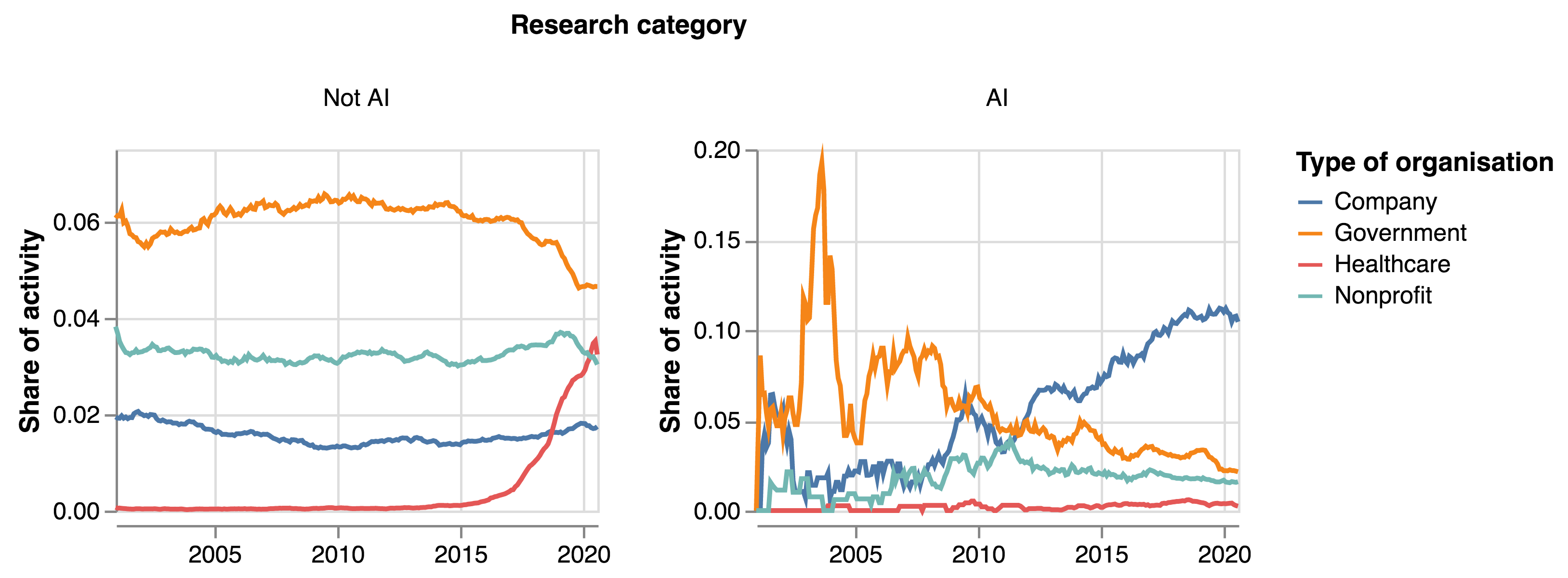}
    \captionsetup{margin=1cm}
    \caption{Evolution of participation of different organisation types outside of AI research (left) and in AI research (right). The scale in the Y axis is different between both panels. Series are smoothed using a 10 month rolling average.}
    \label{fig:org_evol}
\end{figure}

Figure \ref{fig:org_evol} shows the evolution in the shares of research participations by different organisation types excluding academic institutions. The right panel shows that private sector participation in AI research is a recent phenomenon and that it has almost doubled in the last decade. Government institutions used to play a more important role in AI research before 2010, although there were very low volumes of AI research at that point. Healthcare and nonprofit participation in AI research has remained stable in recent years, and below the levels we see outside AI research (which are displayed in the left panel).

\begin{figure}[ht!]
    \centering
    \includegraphics[width=0.9\textwidth]{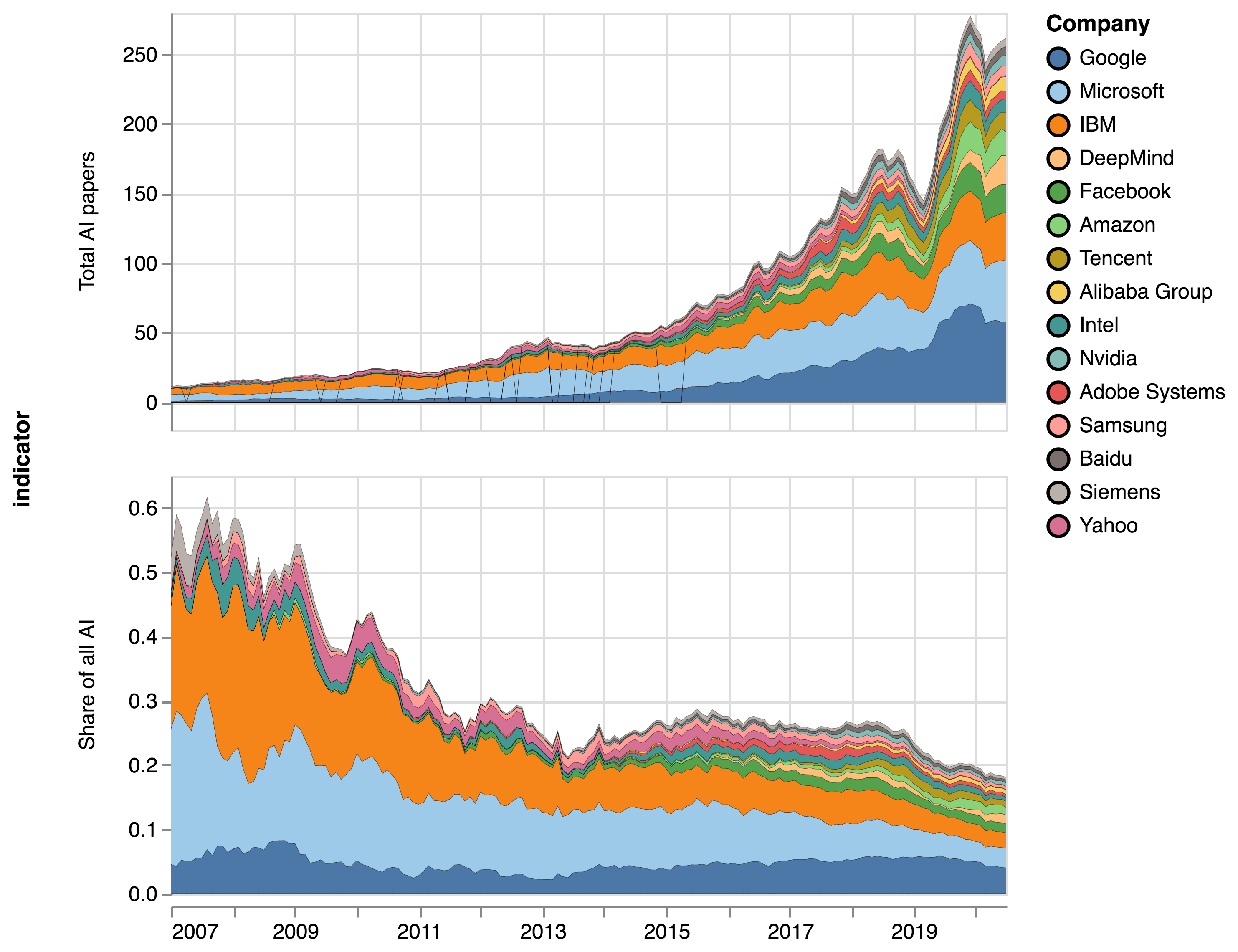}
    \captionsetup{margin=1cm}
    \caption{The top panel presents the evolution in the number of AI papers involving the top 15 companies by total levels of activity in the AI arXiv corpus. The bottom panel presents the evolution of the share of AI papers involving those same companies. Both series are smoothed using an 8-month rolling average.}
    \label{fig:comp_evol}
\end{figure}

Figure \ref{fig:comp_evol} shows levels of AI activity involving the 15 companies most active in the corpus. Our results confirm the idea that technology and internet companies are actively involved in AI research: the top 5 companies in our list are US technology companies - Google, Microsoft, IBM, DeepMind and Facebook. While Google, Microsoft and IBM have been active in AI research since the early 2000s, DeepMind and Facebook only started gaining prominence since the mid 2010s. There are several Chinese internet companies (Tencent, Alibaba and Baidu) in our data, as well as two European manufacturing businesses (Bosch and Siemens) and semiconductor manufacturers Intel and Nvidia, capturing research around AI's hardware components.

When we focus on the share of AI research involving these companies in the bottom panel of Figure \ref{fig:comp_evol}, we find, perhaps surprisingly, that their share of AI activity has \textit{declined} since the mid 2000s, when almost 60\% of AI papers involved them, with particularly strong levels of participation by Microsoft, IBM, Google and Intel. It is worth pointing out that the overall levels of AI research were very low at that point (see top panel). When we consider more recent years when the levels of AI research in arXiv start increasing, we find an initial increase in the share of AI research involving these companies, reaching around 25\% of all AI research between 2015 and 2019. Since 2019 this share has been declining although this is driven by rapid growth in the volume of AI research in wider arXiv rather than a decline in AI publication levels in these companies. In the first months of 2020, almost two in ten AI research papers in arXiv involved one of these private sector companies.

\subsubsection{Comparing the thematic diversity of private and public AI research}

We move on to compare the thematic diversity of the corpora of public and private AI research: do we find, consistent with the claim we developed during the literature review, that private sector companies have a  narrower research profile than those in the public sector?

% One potential mechanism for this kind of decline put forward in the directed technological change literature (and consistent with some of recent trends in AI research outlined in subsection \ref{subsec:lit_private}) is the presence of organisations with strong incentives to focus on those technologies that perform better presently at the expense of alternatives that would preserve technological diversity. We would expect those incentives to be stronger in the private sector than in academia and the public sector. Here we explore this question empirically. 

%  and map AI research organisations in a semantic space whose structure could shed some light on the state and recent evolution of thematic diversity in AI research. 

\begin{figure}[h!]
    \centering
    \includegraphics[width=0.9\textwidth]{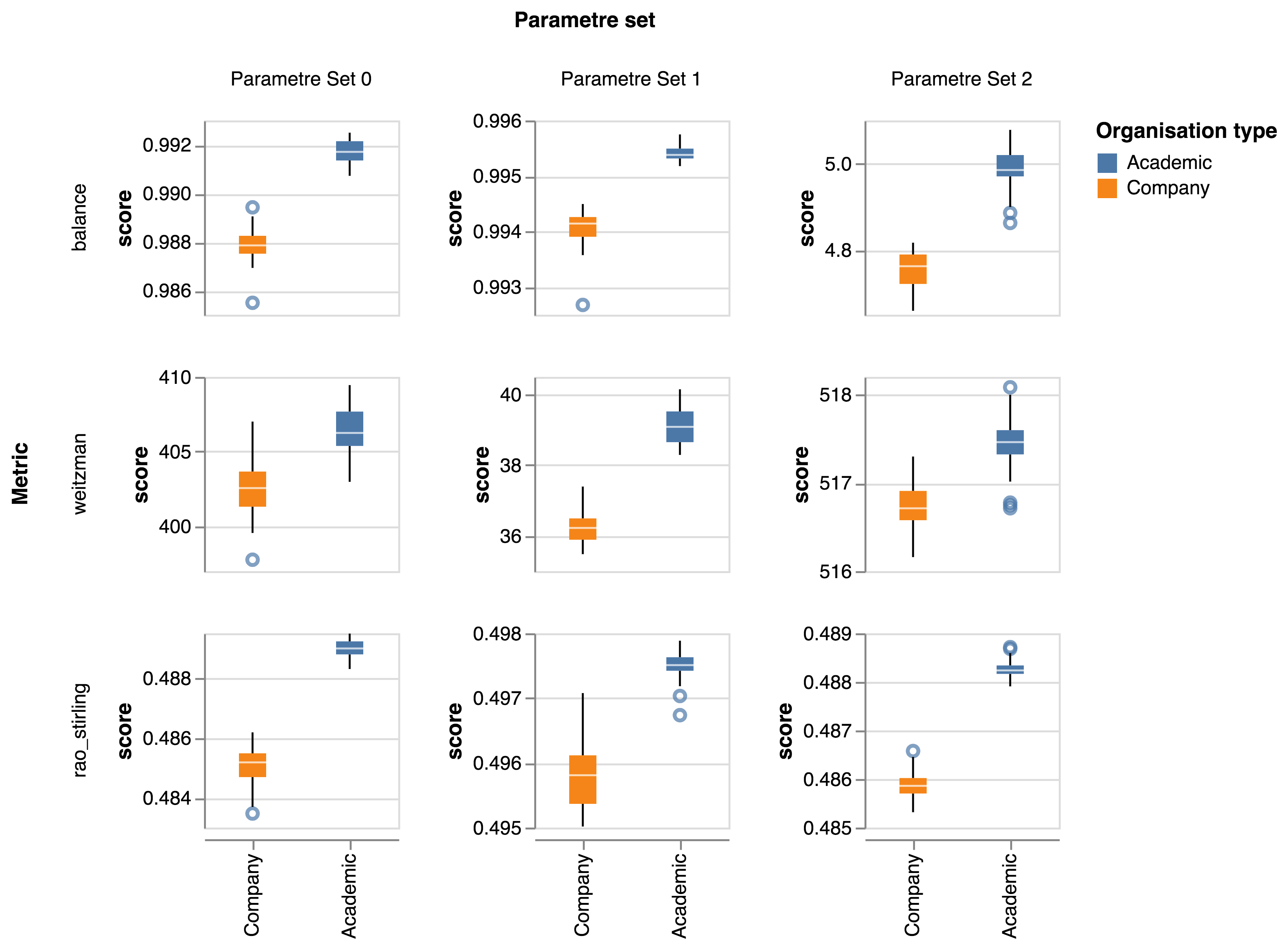}
    \captionsetup{margin=1cm}
    \caption{Thematic diversity in AI research after 2015 and involving / excluding private sector companies according to different metrics of diversity and parametre sets, calculated in samples of 1000 articles drawn from each sub-corpora, with 30 runs per comparison. The scale in the Y axis is different between charts}
    \label{fig:div_comparison_sample}
\end{figure}

Figure \ref{fig:div_comparison_all} compares thematic diversity in the corpus of AI research involving private sector companies and the rest of the corpus. We seek to adjust for differences in the size of the corpora which may confound the results by drawing random samples of 1000 articles from each of our sub-corpora (research involving private sector and non private sector organisations) and use them to calculate thematic diversity with our three metrics and parametre sets. Figure \ref{fig:div_comparison_sample} compares the mean scores for these metrics and parametre sets with 30 random sample draws per metric / parametre set and sub-corpus. We find that thematic diversity in private AI research is consistently lower for all metrics and parametre sets.

\subsubsection{Modelling thematic diversity at the organisation level}

We move on to focus on the drivers of thematic diversity at the organisational level using a linear regression framework. This allows us to build a more finely grained understanding of the micro-dynamics of field level diversity, and to account for heterogeneity of different organisation types.

Our model setup is thus:

\begin{equation}
    d_{i,m,p,y} = \alpha + \beta_1is\_comp_i + \beta_2log(article\_n_{i,y}) + \beta_3y + \epsilon_i 
\end{equation}

Where $d_{i,m,p,y}$ is thematic diversity in organisation $i$ according to metric $m$ and parametre set $p$ in year $y$. $is\_comp$ is a dummy capturing whether $i$ is a company or not, and $log(article\_n_{i,y})$ is the number of articles published by the organisation in $y$ (logged). We include year fixed effects and, in one of the specification of the model, to capture unobservable sources of heterogeneity between organisations. 

We are specially interested in the coefficients for $\beta_1$, the estimate of the link between an organisation type and its thematic diversity after we control for other important factors, in particular its levels of research activity, which could impact independently on its thematic diversity.

We focus our analysis on 188 organisations with at least 75 AI publications in the data, and in the last three years (2018,2019 and 2020).

Table \ref{tab:reg_table} presents our regression results.

\begin{table}[!htbp]
\small
    \centering
    \captionsetup{margin=1cm}
\begin{tabular}{lllllll}
\toprule
       Balance &        1 &        1 &        2 &        2 &        3 &        3 \\
\midrule
 Company index &   -0.07* &    -0.3* &   -0.13* &   -0.43* &   -0.08* &  -0.45** \\
               &  (-0.99) &   (-0.8) &  (-1.79) &  (-1.33) &  (-1.67) &  (-2.25) \\
  Number of papers (log) &  0.86*** &  1.19*** &   1.0*** &  1.51*** &  1.17*** &  1.09*** \\
               &  (12.39) &   (9.18) &  (12.98) &   (9.48) &  (55.26) &  (22.29) \\
          Year &    0.02* &    0.01* &    0.03* &     0.0* &    0.03* &   0.04** \\
               &   (0.52) &   (0.14) &   (0.75) &    (0.1) &   (1.49) &    (2.0) \\
\midrule
         $R^2$ &     0.45 &     0.71 &     0.61 &     0.82 &     0.84 &     0.92 \\
             $N$ &      564 &      564 &      564 &      564 &      564 &      564 \\
 Fixed Effects &       No &      Yes &       No &      Yes &       No &      Yes \\
\bottomrule
\\
\end{tabular}

\begin{tabular}{lllllll}
\toprule
      Weitzman &        1 &        1 &        2 &        2 &        3 &         3 \\
\midrule
 Company index &    0.04* &    0.09* &   -0.01* &    0.04* &  -0.1*** &  -0.23*** \\
               &   (0.69) &   (0.89) &  (-0.15) &   (0.43) &  (-4.41) &   (-2.84) \\
  Number of papers (log) &  1.22*** &  0.86*** &  1.24*** &  0.96*** &  1.26*** &   1.08*** \\
               &  (39.27) &  (18.89) &  (44.25) &  (24.75) &  (67.08) &   (30.12) \\
          Year &   -0.01* &    0.01* &   -0.01* &    0.01* &    0.01* &    0.02** \\
               &  (-0.57) &   (0.89) &   (-0.6) &   (0.74) &   (0.96) &    (2.27) \\
\midrule
         $R^2$ &     0.91 &     0.98 &     0.94 &     0.98 &     0.97 &      0.99 \\
             $N$ &      564 &      564 &      564 &      564 &      564 &       564 \\
 Fixed Effects &       No &      Yes &       No &      Yes &       No &       Yes \\
\bottomrule
\\
\end{tabular}

\begin{tabular}{lllllll}
\toprule
  Rao-Stirling &        0 &        0 &        1 &        1 &         2 &        2 \\
\midrule
 Company index &   -0.09* &   -0.26* &   -0.02* &   -0.12* &  -0.31*** &  -0.72** \\
               &  (-1.28) &  (-0.69) &  (-0.32) &  (-0.28) &   (-3.66) &  (-2.22) \\
  Number of papers (log) &  0.85*** &  1.12*** &  0.72*** &  0.99*** &   0.99*** &  1.37*** \\
               &  (12.75) &   (8.89) &   (9.56) &   (6.71) &   (13.26) &    (9.3) \\
          Year &    0.02* &     0.0* &    0.01* &    -0.0* &     0.04* &    0.02* \\
               &   (0.39) &   (0.08) &   (0.19) &   (-0.1) &    (1.02) &   (0.62) \\
\midrule
         $R^2$ &     0.44 &     0.72 &     0.32 &     0.64 &      0.58 &     0.82 \\
             $N$ &      564 &      564 &      564 &      564 &       564 &      564 \\
 Fixed Effects &       No &      Yes &       No &      Yes &        No &      Yes \\
\bottomrule
\\
\end{tabular}
    \caption{Regression results for various diversity metrics (Balance in top table, Weitzman in middle table, Rao-Stirling in bottom table) and parametre sets (see columns). t-values in parentheses. *** p $<$ 0.01, ** p $<$ 0.05, * p $<$ 0.1.}
    \label{tab:reg_table}
\end{table}

The coefficient $\beta_1$ is negative for all balance and Rao-Stirling metrics of diversity. Interestingly, the effect gains strength and significance when we introduce organisation fixed effects, suggesting that there is heterogeneity inside organisation types. We explore this result in further detail below by looking at the fixed effects of individual organisations in the data. The association between company status and Weitzman diversity are also generally negative but weaker, and in one case (parametre set 1) are positive. This could be explained by the differences between the Weitzman indicator and the other metrics of diversity that we mentioned previously.

In all cases, $\beta_2$, the coefficient for the link between number of papers and diversity metrics is positive, consistent with the idea that broader corpora and larger organisations are, other things equal, able to maintain more thematically diverse research profiles.

The coefficient $\beta_3$ for the link between time and thematic diversity is generally positive, suggesting an increase in thematic diversity over time inside organisations. This could still lead to an stagnation or decline in thematic diversity in the aggregate if there is homogeneity in the topics being pursued by different organisations and/or the composition of the field changes in a way that gives more importance to the activities of less thematically diverse organisations.

\begin{figure}[]
    \centering
    \includegraphics[width=0.9\textwidth]{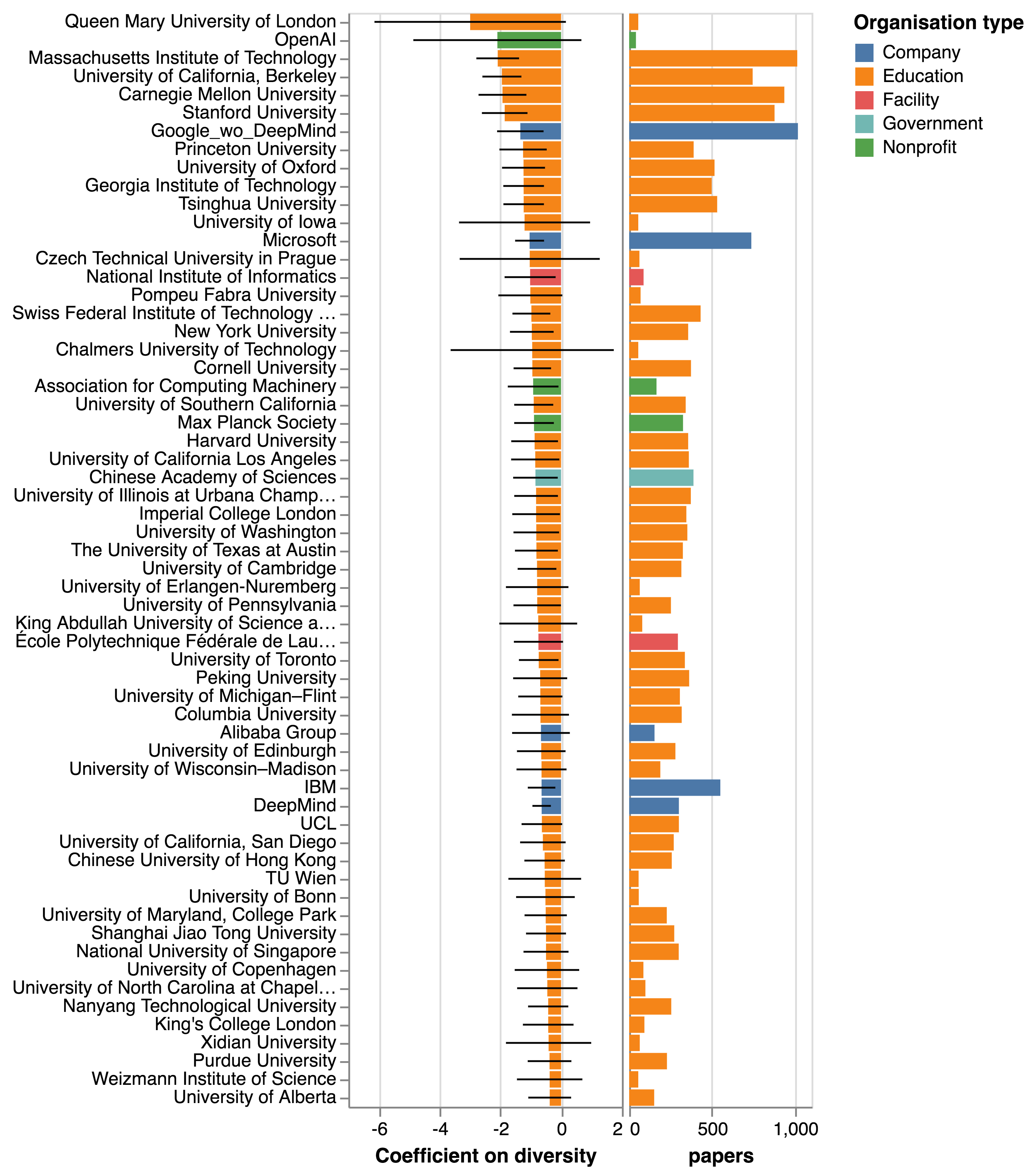}
    \captionsetup{margin=1cm}
    \caption{The right panel shows fixed effects and confidence intervals for the coefficient of top organisations in the Rao-Stirling specification with parametre set 2. The left bar shows the number of AI publications by organisations in the period. The colour of the bar shows the organisation type according to GRID.}
    \label{fig:org_fixed}
\end{figure}

Figure \ref{fig:org_fixed} presents the coefficients for the organisation fixed effects on the Rao-Stirling diversity metric with parametre set 2 in the left panel, and number of papers involving the organisation in the right panel.\footnote{As shown throughout, the Rao-Stirling diversity results are very similar for different parametre sets, an to the balance metrics - we choose parameter set 2 because it has a better goodness of fit than the alternative specifications.} Our fixed effects capture an organisation's diversity compared to the rest of the population after adjusting for its logged publication levels, publication year and its organisation type (whether it is a company or not). The group of most thematically narrow organisations in the data comprises some of the largest and most prestigious US institutions in AI research (MIT, Carnegie Mellon, University of California Berkeley and Stanford) as well as Google and Microsoft.

Although we advise caution in the interpretation of these results (the confidence intervals for most fixed effects are very broad), these estimates suggest that even though, on average, educational institutions tend to be more thematically diverse than private companies, there are some important differences in both groups. Some of the most prestigious and active US universities seem to be thematically narrower than smaller academic institutions in other countries.

\subsubsection{The thematic profile of private sector companies in AI research}
\label{subsec:priv_spec}

Our analysis above suggests that some private sector companies tend to be thematically narrower than other institutions in the corpus. What topics are they focusing on?

\begin{figure}[h!]
    \centering
    \captionsetup{margin=1cm}
    \includegraphics[width=0.9\textwidth]{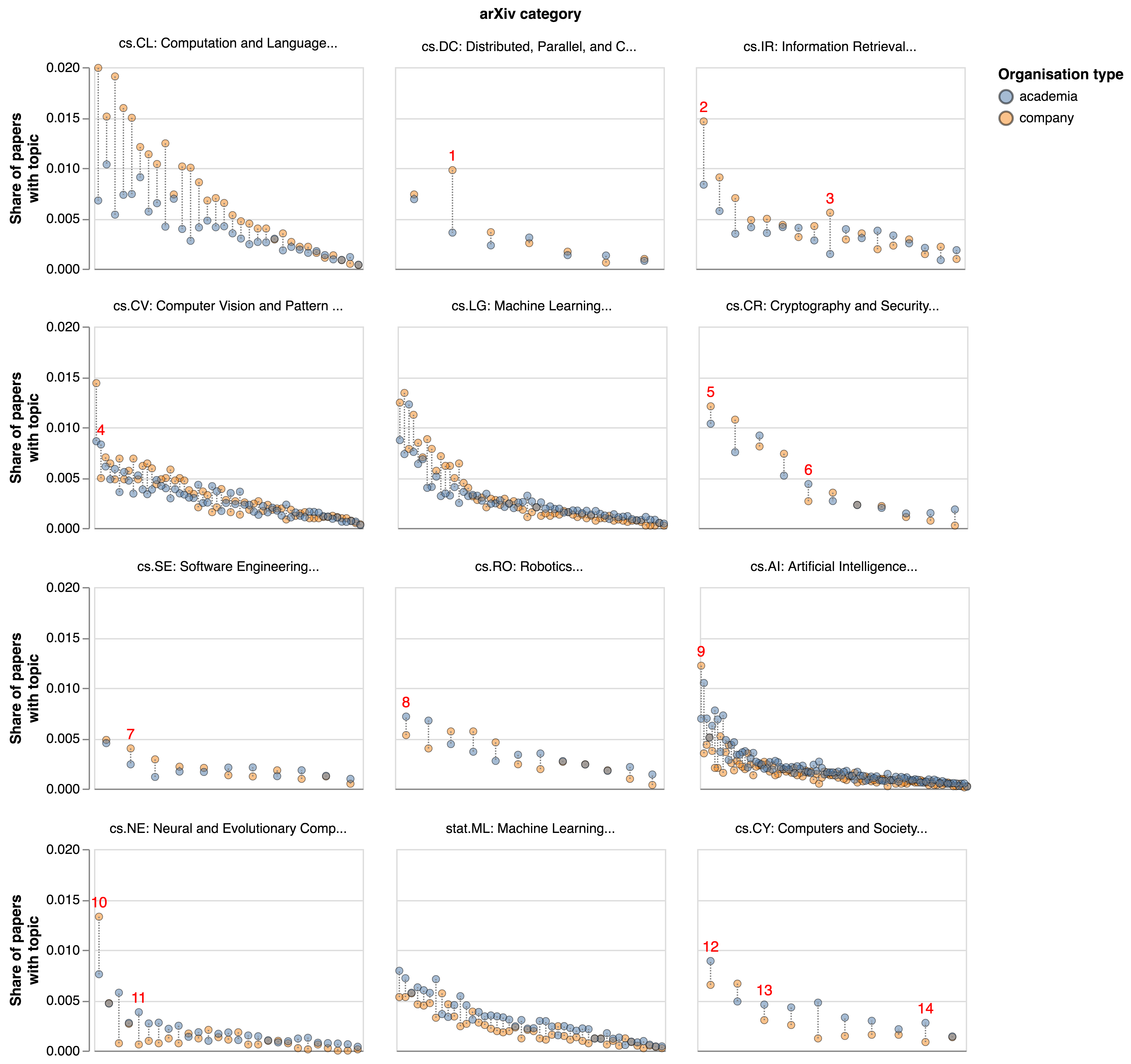}
    \caption{Each scatter presents, for each arXiv category, the share of AI research involving private sector organisations (orange point) and public sector organisations (blue point) with a topic. Topics are assigned to arXiv categories following the approach presented in subsection \ref{subsec:ai_evol}, and sorted by their overall importance in the category. The categories are sorted from left to right and top to bottom based on the mean difference of shares of activity in private sector AI research and public sector AI research for all topics in the category. We have highlighted in red some notable topics (see table \ref{tab:topic_examples} for their labels)}
    \label{fig:type_topic_comparison}
\end{figure}

To answer this question, we have compared the share of AI articles involving private sector institutions with a topic with the share of AI articles that do not involve private sector institutions. This is a proxy for the levels of relative specialisation (or over-representation) of private sector AI research in particular topics. We present the results in Figure \ref{fig:type_topic_comparison}, and some notable examples in Table \ref{tab:topic_examples}.

We find that private sector companies tend to be more specialised in arXiv categories such as \textit{cs.CL} (Computer Language), \textit{cs.DC} (Distributed and parallel computing), \textit{cs.IR} (information retrieval) and \textit{cs.CV} (computer vision). They are less focused on \textit{cs.AI} (Artificial Intelligence concentrating on symbolic techniques), \textit{cs.NE} (Neural and evolutionary computing), \textit{Stat:ML} (statistical machine learning) and \textit{cs.CY} (Computers and Society). This is consistent with the idea that private sector companies focus on applications of AI enabled by deep learning and in research to advance the computational infrastructure required by large-scale, safe deep learning research. They are less focused on AI techniques outside of deep learning, and on wider AI application and implications. 

Table \ref{tab:top_examples} presents some illustrative topics. Private sector AI research specialises in topics related to recommendation systems and advertising (keys 2 and 3), and research about hardware that complements deep learning, such as graphical processing unit optimisation (key 1). In less applied categories such as \textit{cs.AI} or \textit{cs.NE}, private sector companies focus on techniques allied to the deep learning design such as reinforcement learning or recurrent neural networks. Companies are also over-represented in a topic that mentions code and GitHub (a widely used code sharing repository), suggesting that private sector researchers are more likely to release open source code for others to adopt and build on, encouraging the adoption of the techniques that they develop.

Academic researchers tend to be relatively specialised in AI applications in health, including analyses of MRI scans (key 4), healthcare systems (key 12) and infections and pandemics (key 14), as well as research that considers the ethical and legal implications of AI. 

\begin{table}[!htbp]
\small
    \centering
    \captionsetup{margin=1cm}
    \begin{tabular}{p{0.3in} p{3in} p{1in} p{1in}}
    \toprule
    Key & Topic label & arXiv category & Specialised
\\ 
\midrule
1 & optimizations gpu gpus tensorflow cpu & cs.DC & \textbf{Company}
\\ \\
2 & items recommendation recommendations item recommender systems & cs.IR & \textbf{Company}
\\ \\
3 & conversion revenue ads click advertising & cs.IR & \textbf{Company}
\\ \\
4 & registration medical imaging mri scans volumes & cs.CV & Academia
\\ \\
5 & adversarial adversarial examples adversarial training adversarial attacks adversarial perturbations & cs.CR & \textbf{Company}
\\ \\
6 & security iot secure protection vulnerabilities & cs.CR & Academia
\\ \\
7 & code source code bugs apis github & cs.SE & \textbf{Company}
\\ \\
8 & driving vehicle vehicles autonomous driving road & cs.RO & Academia
\\ \\
9 & reinforcement learning deep reinforcement learning model free q learning reinforcement learning algorithms & cs.AI & \textbf{Company}
\\ \\
10 & recurrent lstm rnn recurrent neural networks rnns & cs.NE & \textbf{Company}
\\ \\
11 & swarm pso abc particle swarm optimization metaheuristic & cs.NE & Academia
\\ \\
12 & clinical patients medical patient healthcare & cs.CY & Academia
\\ \\
13 & law society legal ethical stakeholders & cs.CY & Academia
\\ \\
14 & infection pandemic epidemic infected virus & cs.CY & Academia
\\ \\
    \bottomrule
    \end{tabular}
    \caption{Legend for topics in Figure \ref{fig:type_topic_comparison}. The key is the key for the topic in that figure, the topic label its name, the arXiv category is the category where the topic is most salient, and specialised shows whether companies or academic institutions are most specialised in the topic.}
    \label{tab:top_examples}
\end{table}

\subsection{The influence of private sector organisations in AI research}
\label{subsec:influence}

One potential interpretation of our analysis so far is that increasing participation by private companies with narrow research profiles in AI research could be narrowing the field by altering its composition. However, and as pointed out in previous sections, private companies comprise less than 20\% of AI research participations and their share of publications has not increased substantially in recent years. In this subsection, we explore two other mechanisms through which private AI research may be influencing indirectly the trajectory in AI in a way that makes it narrower: by influencing what other researchers do (imperfectly proxied via citations), and through collaboration with influential institutions.

\subsubsection{Citations to private AI research}
 Articles involving private companies could be having an out-sized influence in the evolution of AI research despite being in the minority of our corpus. To assess this, we compare the citation counts of articles involving / not involving private companies. As table \ref{tab:cit_descr} shows, these articles receive around twice as many citations as articles without private companies. The median number of citations received by private companies is higher, suggesting that this result is not driven by outlier articles with large numbers of citations. 

\begin{table}[]
    \centering
\begin{tabular}{lrrrr}
\toprule
 Article category &       Mean &  Median &         std &      Max \\
\midrule
   Not involving company &  13.583 &     0.0 &  106.256 &  10683 \\
    Involving company &  26.498 &     1.0 &  218.754 &   9729 \\
\bottomrule
\end{tabular}
    \caption{Descriptive statistics for citation counts depending on article category }
    \label{tab:cit_descr}
\end{table}

\begin{figure}[h!]
    \centering
    \captionsetup{margin=1cm}
    \includegraphics[width=0.9\textwidth]{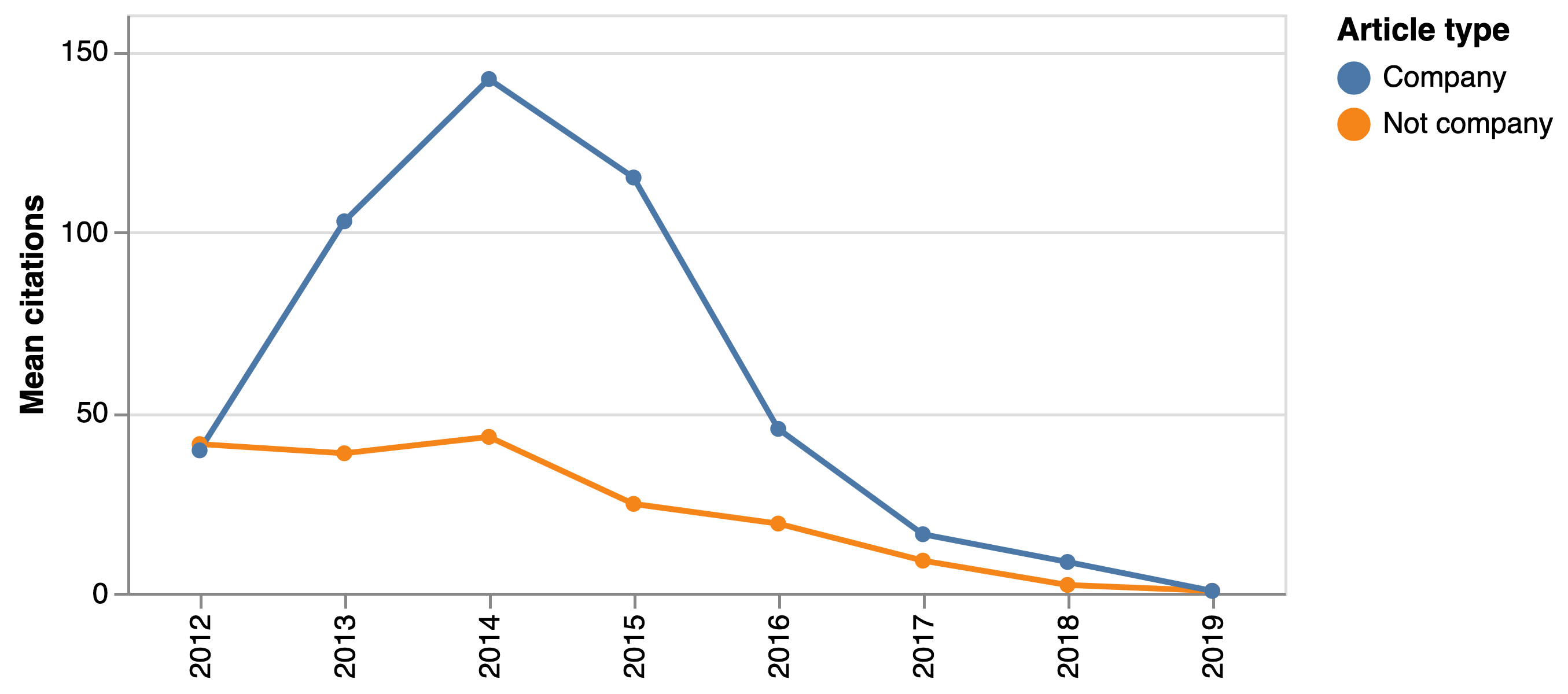}
    \caption{Mean citation count for articles involving / not involving private companies}
    \label{fig:cit_comp}
\end{figure}

In Figure \ref{fig:cit_comp} we compare the evolution of mean citation counts for papers involving / not involving private companies. It shows that corporate AI research gained influence during the first half of the 2010s, coinciding with the arrival of deep learning. The differences in mean citation counts dwindles in recent years although this is probably caused by the fact that newer papers have had less time to accrue citations.

Some of the differences between private companies and public institutions could be explained by differences in their thematic focus that we explored in subsection \ref{subsec:priv_spec}: if private companies specialise in popular topics related to deep learning, then this might explain why they receive more citations. In order to account for this, we create a model where we regress the number of citations received by a paper on other relevant variables. The model is specified as:

\begin{equation}
    cit_i = \alpha + \beta_1is\_comp_i + \beta_2num\_insts + \beta_3year + \beta_4topic\_comp_i + \epsilon
\end{equation}

Here, $cit_i$ is the number of citations received by an article, $is\_comp$ captures whether the paper involved at least on company, $num\_insts$ the number of institutions involved in the paper, $year$ the year when the paper was published, and $topic\_comp$ a vector capturing the topical composition of the paper. We include this variable with the goals of controlling for differences between the thematic focus of private sector researchers and public sector researchers. Since citation count data is highly skewed, we fit the model with a regularised Poisson regression. We also decompose the 558-dimensional outputs of the topSBM model into a smaller number of components using principal component analysis with a variable number of components to check the robustness of our analysis and the effect of including progressively more granular descriptions of the topical composition of an article in the citation model. We present the results in table \ref{tab:cit_reg}.

\begin{table}[!h] \centering \small
\begin{tabular}{@{\extracolsep{5pt}}lcccc}
\\[-1.8ex]\hline
\hline \\[-1.8ex]
& \multicolumn{4}{c}{\textit{Dependent variable: $cit\_count$}} \
\cr \hline
\\[-1.8ex] & \multicolumn{1}{c}{No topics} & \multicolumn{1}{c}{PCA 10} & \multicolumn{1}{c}{PCA 50} & \multicolumn{1}{c}{PCA 100}  \\
\\[-1.8ex] & (1) & (2) & (3) & (4) \\
\hline \\[-1.8ex]
 Constant & 331.739$^{***}$ & 358.131$^{***}$ & 390.943$^{***}$ & 395.528$^{***}$ \\
  & (6.570) & (7.749) & (9.682) & (9.450) \\
 is\_comp & 0.898$^{***}$ & 0.835$^{***}$ & 0.757$^{***}$ & 0.761$^{***}$ \\
  & (0.099) & (0.098) & (0.096) & (0.097) \\
 num\_auth & 0.019$^{***}$ & 0.019$^{***}$ & 0.022$^{***}$ & 0.025$^{***}$ \\
  & (0.006) & (0.006) & (0.007) & (0.007) \\
 year & -0.163$^{***}$ & -0.176$^{***}$ & -0.193$^{***}$ & -0.195$^{***}$ \\
  & (0.003) & (0.004) & (0.005) & (0.005) \\
\hline \\[-1.8ex]
Pseudo-$R^2$ & 0.198 & 0.228 & 0.285 & 0.313 \\
 Observations & 57,924 & 56,456 & 56,456 & 56,456 \\
\hline
\hline \\[-1.8ex]
\end{tabular}
\captionsetup{margin=1cm}
\caption{Results of Poisson regression of citation counts without including an article's topical composition (column 1) and including a PCA decomposition of topic model with 10, 50 and 100 components (other columns). Robust standard errors in parentheses. {$^{*}$p$<$0.1; $^{**}$p$<$0.05; $^{***}$p$<$0.01}}
\label{tab:cit_reg}
\end{table}

We find that articles involving private companies tend to receive more citations after accounting for the year they were published, the number of institutions involved in the paper and their topical composition. Including the thematic composition of a paper reduces the coefficient on the $is\_comp$ variable, consistent with the idea that private companies specialise in more popular research topics, and increases a model's goodness of fit. 

The conclusion of this analysis is that AI research undertaken by private companies is particularly influential, which could steer the evolution of the field in a direction that resembles their narrow research profiles.

\subsubsection{Private sector collaboration patterns}

Private companies could also be influencing AI research through their collaborations with other institutions. This could narrow thematic diversity if, for example, they provide access to skills, data and computational infrastructure that helps collaborators adopt deep learning techniques along the lines discussed in \citep{wahmed2020}. In order to explore this issue, we analyse the collaboration patterns private and public researchers, based on their co-authorship patterns in arXiv AI papers.

We find that 60\% of the papers involving private companies also include at least one educational institutions, suggesting significant levels of industry-academia collaboration in AI research. We then rank educational institutions based on their level of collaboration with private companies and calculate the mean number of citations that all their AI research  - table \ref{tab:ai_collab} presents the results.

\begin{table}[]
    \centering
    \small
\begin{tabular}{p{1.1in} p{1.1in} p{3.5in}}
%\begin{tabular}{cll}
\toprule
Private collaboration rank &  Mean Citation count &                                                                                                                                                       Top 5 organisations \\
\midrule
       1-10 &          23.411 &                         Carnegie Mellon University, Stanford University, Massachusetts Institute of Technology, University of California, Berkeley, Georgia Institute of Technology \\
      11-50 &          21.370 &                                                        Peking University, The University of Texas at Austin, University of Michigan–Flint, Columbia University, New York University \\
     50-100 &          13.097 &  Hebrew University of Jerusalem, Ludwig Maximilian University of Munich, Hong Kong University of Science and Technology, Indonesian Institute of Sciences, Arizona State University \\
    100-500 &          13.272 &                                Harbin Institute of Technology, Centre national de la recherche scientifique, Rice University, Polytechnique Montréal, Royal Institute of Technology \\
  Above 500 &          10.286 &                                                          Ohio University, Sao Paulo State University, Saint Mary's University, South China Normal University, University of Münster \\
\bottomrule
\end{tabular}

    \caption{Mean citation counts and top institutions in different rankings of the distribution of research collaboration with private companies.}
    \label{tab:ai_collab}
\end{table}

We find a monotonic decrease in the mean citation counts of organisations that collaborate with AI researchers as we go down the collaboration rankings, suggesting that private companies tend to work together with more influential institutions. The top collaborators with private AI researchers are elite institutions in the US which, as we showed in our organisational analysis of thematic diversity, tend to have relatively narrow research profiles.

\section{Conclusion} \label{sec:conclusion}

\subsection{Discussion of results}

We have studied the thematic diversity of AI research in the arXiv pre-print corpus using a variety of metrics, parametre sets and approaches with the goal of addressing three research questions: Is the thematic diversity of AI research increasing or declining? How does the thematic diversity of private sector organisations compare with those in academia and the public sector? How influential is private sector AI research?

% We do this motivated by the innovation studies and economics of science, technology and innovation literatures, which have identified a set of processes that may lead to a loss of diversity in a technology landscape and its dominance by technologies that are (or are eventually found to be) inferior. Recent trends in AI research that we also document suggest the presence of such trends as powerful deep learning systems build momentum despite some misgivings about their limitations, the field becomes increasingly dominated by private sector companies focused on technologies that complement their assets and capabilities, and processes of institutional isomorphism lead to a convergence in the behaviours and profiles of organisations in the field.

Regarding the first question, our findings are broadly consistent with the idea of a narrowing of AI research: after an initial period where thematic diversity in AI research increased as deep learning techniques emerged and started to be deployed in a variety of settings, thematic diversity has stagnated and perhaps even started to decline in recent years. This finding is robust to changes in the approach we use to identify AI papers, topic modelling strategy, and the creation of a time-adjusted corpus to remove biases caused by the concentration of AI research in recent years.

% We see greater concentration of research activity in popular research techniques (topics) and a decline in the disparity of these techniques.

Regarding the second question, our comparison of the thematic diversity of private sector organisations with other participants in AI research suggests that the former are more narrowly focused on a specialist set of techniques related to the digital economy such as computer language, computer vision and information retrieval. These companies also specialise in research topics about hardware and infrastructure to scale up deep learning systems. Companies are less focused on health applications of AI and ethical and legal considerations. We also find some evidence of institutional isomorphism in the field: elite universities in the US that tend to collaborate actively with the private sector have comparatively narrow research profiles. This suggests that smaller academic institutions might be playing a role helping to preserve thematic diversity in AI research.

Our analysis of citation and collaboration patterns to address the third research question shows that private AI research tends to receive more citations even after we control for its topical focus, and that private companies collaborate actively with the most prestigious academic institutions. We speculate that these collaborations could contribute indirectly to narrow thematic diversity in AI research by influencing follow-on research and the research agendas of key players in AI research in the direction of adopting techniques related to the dominant deep learning design. Establishing this more precisely will be a key question for future work.

\subsection{Limitations and issues for further research}

Throughout the paper we have assumed that there is an intrinsic value in technological / thematic diversity drawing on theoretical arguments developed in subsection \ref{subsec:diversity_value} the history of AI, where previous efforts to preserve technological diversity provided the foundation for subsequent advances (including the back-propagation algorithm which, as our epigraph notes, for decades was \textit{`cool math that didn't accomplish anything'} but eventually provided a key building block for the deep learning revolution), and by concerns about important limitations in the deep learning trajectory. However, it could also be the case that some thematic diversity in AI research is dysfunctional, reflecting for example scholarly inertia and lack of skills or infrastructure preventing academic researchers from adopting state-of-the-art techniques. Some of our metrics of diversity, by estimating this variable in relative rather than absolute terms, may be underestimating the range of ideas and methods being explored by a growing community of AI researchers.

Future research should set out to quantify the value of thematic diversity in terms of its contributions to resilience, inclusion and creativity that we have articulated in this paper, addressing questions such as: to which extent are novel combinations of ideas helping overcome some of the limitations of the dominant AI design and enabling the development of AI applications in novel domains that are poorly served by state-of-the-art commercial technologies? How does the work of organisations with more diverse research profiles contribute to advances in the field, for example by providing components for combinatorial AI innovations?

A worthwhile expansion of our analysis here would be to imitate Weitzman's strategy for the economic valuation of ecological diversity and try to measure the costs and benefits of preserving diversity in AI research taking into account current levels of activity in different topics and the `minimum viable threshold' below which a research topic becomes unsustainable  \citep{weitzman1993preserve}.

Our analysis does not consider the mechanisms underpinning the co-evolution of organisations and technologies in AI research: for example, we tacitly assume that the prominence and influence of private sector companies in AI research is shaping the trajectory of the field. This assumption needs to be tested empirically with a strong focus on identifying the mechanisms through which this `shaping' takes place. For example, to which extent do we see changes in the thematic focus and thematic diversity of organisations that cite or collaborate actively with private sector organisations, or of researchers who transition from academia into industry? This kind of micro analysis could help to generate evidence about the causal impact of private sector participation in AI research, something that our descriptive analysis does not do.

% This could be done by analysing citation patterns between private and public research, studying labour flows between the private sector and academia, and considering the extent to which competitive dynamics in academia - including races to publish and present research in high profile conferences could be playing a role in the narrowing of AI research. Micro-analyses of the behaviours of individual researchers, organisations and communities and their interconnections could shed light on these questions, with important implications for research funders who may want to preserve thematic diversity in AI research. 

Finally, our analysis is based on a single data source about research pre-prints that we analyse using semantic methods. Going forward it will be important to expand this analysis using other data sources capturing AI development and deployment including peer-reviewed publications, patents, open source software development and business development and diffusion activities. Further analyses of thematic diversity in AI using alternative metrics and operationalisations and measurement approaches such as for example citation patterns along the lines of \cite{frank2019evolution} would help validate the results we have presented here. We have released all the code that we have used in our analysis as well as the underlying data to enable such work. \footnote{The code can be accessed \textbf{link to be added}. A list of links to relevant datasets in Figshare can be accessed \textbf{link to be added}.}

% \footnote{The code can be accessed \href{https://github.com/nestauk/narrowing_ai_research}{\textbf{here}}. A list of links to relevant datasets in Figshare can be accessed \href{https://github.com/nestauk/narrowing_ai_research/blob/master/data/sources.md}{\textbf{here}}.}

\subsection{Policy implications}

We have presented a theoretical, qualitative and quantitative rationale for policies to preserve thematic diversity in AI research. Although we currently lack sufficient evidence about mechanisms to confidently recommend specific interventions to that end, we highlight some available options highlighted elsewhere in the literature and in some cases already starting to be implemented in practice. They include increasing the diversity in the backgrounds and disciplines of the individuals and communities involved in AI research \citep{stathoulopoulos2019gender, acemoglu2011diversity}, reducing the brain drain of academic researchers into industry \citep{hain2020}, ensuring that academic researchers have sufficient computational resources to undertake their work independently from the private sector \citep{ho2021building}, developing new benchmarks of AI system performance that capture the strengths of alternative designs \citep{martinez2021research}, using horizon scanning methods to develop funding agendas that take into account counterfactual trajectories in AI research \citep{whittlestone2021and} and funding mission-oriented work to deploy AI systems in domains that may require new techniques \citep{mazzucato2018mission}.

We conclude by highlighting three important challenges standing on the way of these policies.

First, there is an \textit{incentive problem}. Policymakers face a similar dilemma to AI researchers: they have incentives to support those AI technologies with the greatest present potential because they are more likely to produce economic impacts and help develop competitive AI industries. The notion of an AI global race provides an additional geopolitical push to focus on today's state-of-the-art technologies at the expense of investments in technological diversity that may yield benefits in the future or be captured by other organisations or countries.

Second, there is an \textit{information problem}. Even if they overcame the incentive problem, policymakers would need to develop strategies to identify suitable research lines that help preserve thematic diversity \citep{aghion2009science}. Here, they face an information asymmetry with researchers and businesses that have private information about the quality of their research (including its uniqueness and how it could help to diversify AI research) and are thus able to behave opportunistically to secure funding. Strategies to preserve thematic diversity will tend to focus on ideas and communities outside of the research mainstream that are harder to assess reduces their success prospects, and increases the likelihood of wasting resources and distorting research efforts in unproductive directions.

Third, there is a \textit{scale problem}. Most public research budgets are small compared to the R\&D budgets of the large technology companies \citep{lundvall_chinas_2022}. Targeted efforts to steer the trajectory of AI research and increase its diversity are unlikely to have significant impacts unless they are complemented by allied policies to steer technology development and adoption in the private sector. This could include regulatory interventions that penalise algorithmic failures, the environmental costs of AI systems and heavy usage of personal data, thus giving the private sector incentives to develop and adopt alternative techniques \citep{acemoglu2011diversity}.

Recent trends in AI research and the evidence that we have provided suggest that research funders and other stakeholders in the AI policy ecosystem should explore strategies to overcome these challenges and mitigate the risk of a decline in AI's technological diversity, preserving spaces for public interest AI R\&D, developing AI techniques that can be applied in sector with less data and high-stakes decisions, and patiently investing in runner-up AI techniques that might under-perform today but could play an important role in future AI breakthroughs. We hope that the methods and metrics that we have developed in this paper can play a part in informing this policy effort.

\bibliography{references}

\section*{Technical annex}

\subsection*{Detailed procedure to identify AI papers outside of core categories}

We take the following steps to identify AI papers outside of the core categories in our corpus.

\begin{enumerate}
    \item \textit{Preprocess text in the corpus of arXiv abstracts}: This includes lower-casing, removing symbols, numbers and commonly used stop-words and tokenising the abstracts, and combining commonly co-occurring tokens into bi-grams (e.g. 'machine learning') and tri-grams ('deep neural network').
    \item \textit{Define salient vocabulary in each AI category}: We Identify salient terms in the sub-corpora that belong to each of the AI categories: Given the vocabulary in an AI category, for each n-gram with a frequency below 1,000 occurrences we calculate the ratio of its frequency in the category with its frequency in the broader corpus. We select the top 20 n-grams according to this measure of salience.  
    \item \textit{Expand the salient vocabulary for an AI category}: We expand the list of salient terms in each category with a list of similar terms based on a Word2Vec semantic model trained on the whole corpus \citep{mikolov2013efficient}. This allows us to identify terms that appear in a similar context to salient terms in the AI category.\footnote{We select the top 30 terms by similarity to those in $S_i$ with a similarity score above 0.5.}
    \item \textit{Identify articles with high frequencies of n-grams from the expanded salient vocabulary}. Here, we assume that AI articles outside of an AI core category will have a number of terms related to AI above the average for non-AI core categories, but lower than the average for AI core categories. To operationalise this idea, we count the occurrences of salient and expanded terms in articles outside the core AI categories. Any article with a frequency of such n-grams above a critical value $k$ is incorporated into the AI corpus. We define $k= \frac{(\bar{t}_{\bar{c}} +Fs(t_{\bar{c}})) + \bar{t}_c)}{2}$ where $\bar{t}_{\bar{c}}$ is the mean of salient terms for a category in articles outside it, $s(t_{\bar{c}}))$ is the standard deviation of salient terms for a category outside it, $\bar{t}_c$ is the mean of salient terms for a category inside it, and $F$ is a scaling factor that we select depending on the generality / ambiguity of the salient terms that define a core AI category.
\end{enumerate}

We manually inspect the results with different parametres paying special attention to the scaling factor $F$, and choose a value that reduces the number of false positives (non-AI articles classified as AI) that may bias upwards our measures of diversity. 

Tables \ref{tab3a:cat_terms} and \ref{tab3b:cat_terms} shows, for each core AI category, the scaling factor we use to set our critical threshold (higher values mean that we require a higher frequency of salient words in an abstract before including it in our expanded corpus), the frequency of salient terms inside the corpus $C_i$ and outside $O_i$, and the expanded salient vocabulary. It shows, for example, that the salient vocabulary \textit{cs.AI} and \textit{cs.NE} often include quite generic terms (\texttt{agent}, \texttt{belief}, \texttt{search}) which could produce false positives in our search - this is why we set higher scaling factors for these two categories.\footnote{As an example, with lower scaling factors the expansion of $cs.AI$ created false positives with economics research that also uses the language of actors, utilities and preferences.} We note the strong presence of deep learning related terms in the \textit{cs.NE}, \textit{cs.LG} and \textit{stat.ML} categories and the strong overlap between the vocabularies of \textit{cs.LG}, \textit{stat.ML}, reflecting the fact that articles are often labelled with both categories. Meanwhile, \textit{cs.AI} includes terms more often associated to symbolic approaches to AI that dominated previous eras of AI research.

\begin{landscape}
\begin{table}[!h]
    \centering
    \small
    \begin{tabular}{p{1in} p{0.3in} p{0.4in} p{0.4in} p{5.5in}}
    \toprule
         AI category $i$ & $F_i$ & $\bar{f_i(C_i)}$ & $\bar{f_i(O_i)}$ & Expanded salient vocabulary $X_i$
         \\
         \toprule
         \\
         cs.AI & 2 & 1.4 & 0.1 & humans, strategic, beliefs, reasoning, reinforcement learning, causal inference, actions,  agents ,  rules, experts, deep reinforcement learning, online learning, practitioners, artificial intelligence, multi agent, reward, bayesian, engagement, learners, intention, semantics, meta learning, policies, decision, knowledge base, agents, learning algorithms, programming languages, recommendations, agent, decisions, knowledge, learning, cognition, reinforcement, belief, priority, incentives, learner, active learning, decision making, answer, expert, rewards, exploration, causal, decision tree, planning, policy, perception
         \\ \\
         cs.NE & 2 & 3 & 0.4 &  learning ,  network architectures,  task,  functions,  recurrent neural network ,  supervised learning,  unsupervised learning,  generative adversarial networks gans,  deep reinforcement learning,  deep neural network,  semi supervised learning,  algorithm,  neural network,  online learning ,  convolutional neural network cnn, tasks,  convolutional neural networks,  deep convolutional neural networks,  optimizations,  machine learning algorithms, classification,  multi task learning,  bayesian optimization, classification tasks,  deep learning,  input,  deep neural networks,  solutions,  training,  convolutional neural networks cnns,  search,  evolutionary,  architecture,  deep convolutional neural network,  reinforcement learning,  learning algorithms,  network architecture,  artificial neural networks,  recurrent neural networks,  deep neural networks dnns,  optimization,  networks,  machine learning techniques,  meta learning,  convolutional neural network,  genetic,  algorithms,  problems,  neural networks,  dictionary learning 
         \\ \\
         \bottomrule
             \end{tabular}
    \caption{Expansion statistics and vocabulary for cs.AI (Artificial Intelligence) and cs.NE (Neural and Evolutionary Computing)}
    \label{tab3a:cat_terms}
\end{table}
\begin{table}[!h]
    \small
    \centering
    \begin{tabular}{p{1in} p{0.3in} p{0.4in} p{0.4in} p{5.5in}}
    \toprule
         AI category $i$ & $F_i$ & $\bar{f_i(C_i)}$ & $\bar{f_i(O_i)}$ & Expanded salient vocabulary $X_i$
         \\
         \toprule
         \\
         cs.LG & 1 & 1.16 & 0.03 & learning, interpretability, regret, recurrent neural network, supervised learning, unsupervised learning, active learning, generative adversarial networks gans, deep reinforcement learning, deep neural network, semi supervised learning, sgd, neural network, online learning, convolutional neural network cnn, learning based, machine learning, convolutional neural networks, deep convolutional neural networks, machine learning algorithms, dnn, multi task learning, deep learning, bayesian optimization, transfer learning, classification tasks, adversarial examples, interpretable, deep neural networks, reward, convolutional neural networks cnns, dnns, federated learning, deep convolutional neural network, feature selection, reinforcement learning, learning algorithms, artificial neural networks, deep neural networks dnns, recurrent neural networks, machine learning techniques, meta learning, convolutional neural network, sample complexity, gradient descent, classifiers, neural networks, adversarial training, classification task, dictionary learning
         \\ \\
         stat.ML & 1 & 1.4 & 0.1 & semi supervised learning, architectures, reinforcement learning, learning based, learns, architecture, convolutional neural network, deep reinforcement learning, network architecture, trained, reward, convolutional neural network cnn, training, learned, neural networks, embeddings, convolutional neural networks, deep learning, deep convolutional neural network, learners, neural network, tasks, meta learning, deep neural network, labels, gradients, transfer learning, learning algorithms, artificial neural networks, convolutional neural networks cnns, network trained, learning, deep neural networks dnns, unsupervised learning, multi task, adversarial training, autoencoder, adversarial, train, deep convolutional neural networks, deep neural networks, deep, supervised learning, recurrent neural network, dictionary learning, multi task learning, learnt, gradient descent, generative adversarial networks gans, recurrent neural networks
         \\ \\
         \bottomrule
    \end{tabular}
    \caption{Expansion statistics and vocabulary for cs.LG (Machine Learning) and stat.ML (Machine Learning)}
    \label{tab3b:cat_terms}
\end{table}
\end{landscape}

\subsubsection*{An alternative strategy to identify AI papers}
\label{subsubsec:robust_def}

We use the Semantic Scholar API to extract the citations to and from a random sample of 10,000 papers in our corpus \citep{ammar2018construction}. If our corpus was comprehensive, we would expect it to be self-referential i.e. capture the arXiv articles that most influence / are influenced by it, which we would also expect to be strongly related to AI and therefore relevant for our analysis. 

\begin{figure}[!h]
    \centering
    \includegraphics[width=0.9\textwidth]{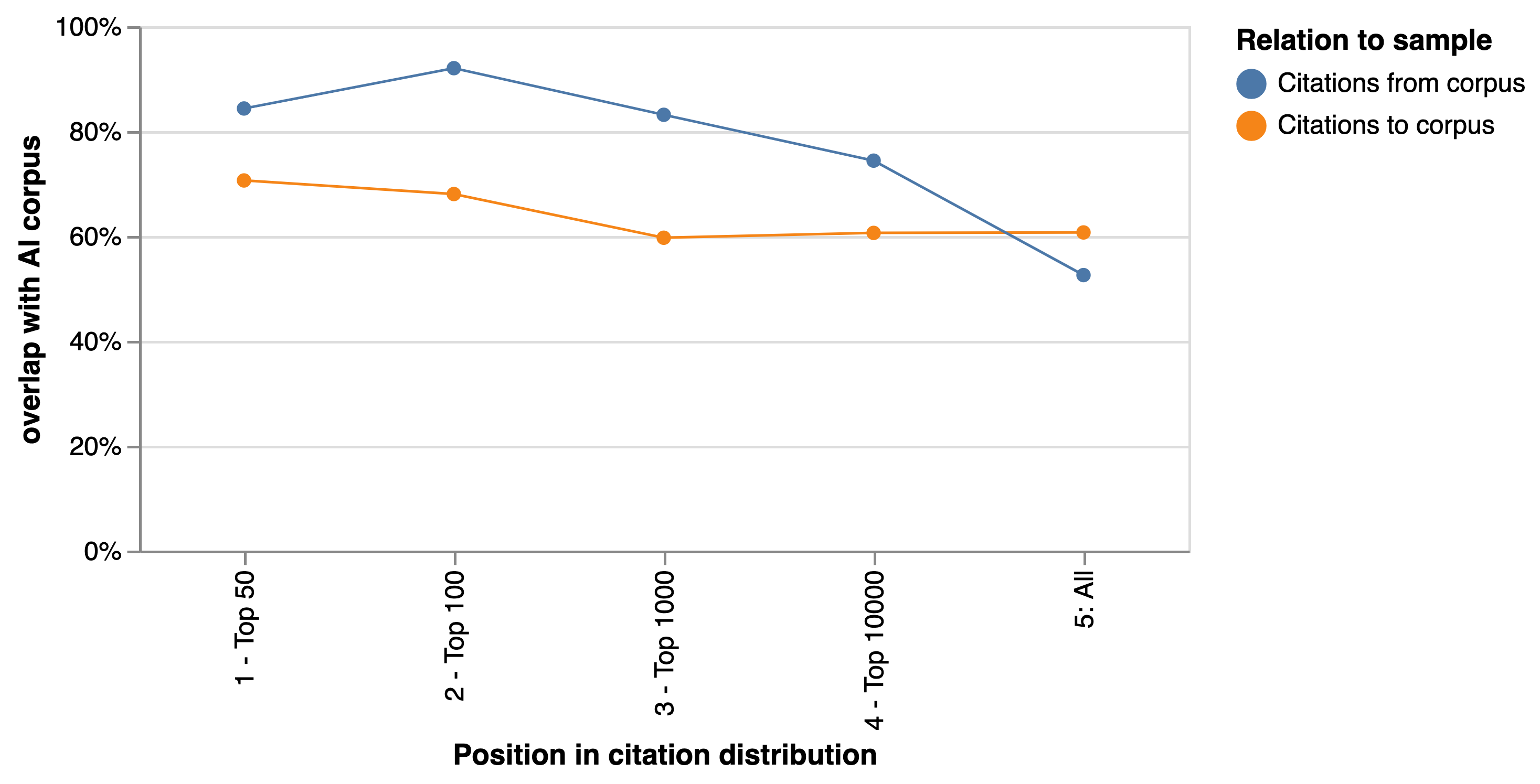}
    \captionsetup{margin=1cm}
    \caption{Proportion of papers in different positions of the citing / cited distribution that are part of our AI corpus.}
    \label{fig_rob:cit_props}
\end{figure}

In total, we extract 72,603 citations from other arXiv articles to this sample corpus (this excludes articles published in 2021 which by definition cannot be part of the original AI corpus) and 360,730 citations from our corpus to other arXiv articles. We rank those articles by the number of times they are cited from / cite our corpus, and visualise, in Figure \ref{fig_rob:cit_props}, the proportion of articles in different positions of the citation / citing distribution that are part of our main AI corpus.

The figure shows that the majority of the papers that cite to / cite from our sample are already part of our corpus. The proportion is higher for those papers that are more influenced by / influential in our sample. We note that the proportion of papers \textit{within} corpus is lower when we consider papers that cite it, raising the possibility that we may be missing papers that apply AI methods in other domains.

In order to evaluate the extent to which this could be creating coverage biases in our topic modelling and diversity analysis, we train a topic model on a combined corpus of papers that cite the AI corpus but are not included in it, papers that are cited from the AI corpus but are not included in it, and a sample of 30,000 papers from the AI corpus, and compare their topic distributions : if these distributions are similar, this would suggest that our corpus is thematically representative of potentially relevant papers outside of it, and that our results would not be altered significantly by their inclusion.\footnote{We use a Latent Dirichlet Model notwithstanding the limitations we highlighted in order to speed up implementation with the \texttt{tomotopy} Python package. We set the number of topics to 300 without any tuning because we are interested in comparing the distribution of topics between different sub-corpora, rather than using the modelling outputs in subsequent analyses.}

\begin{figure}[]
    \centering
    \includegraphics[width=0.9\textwidth]{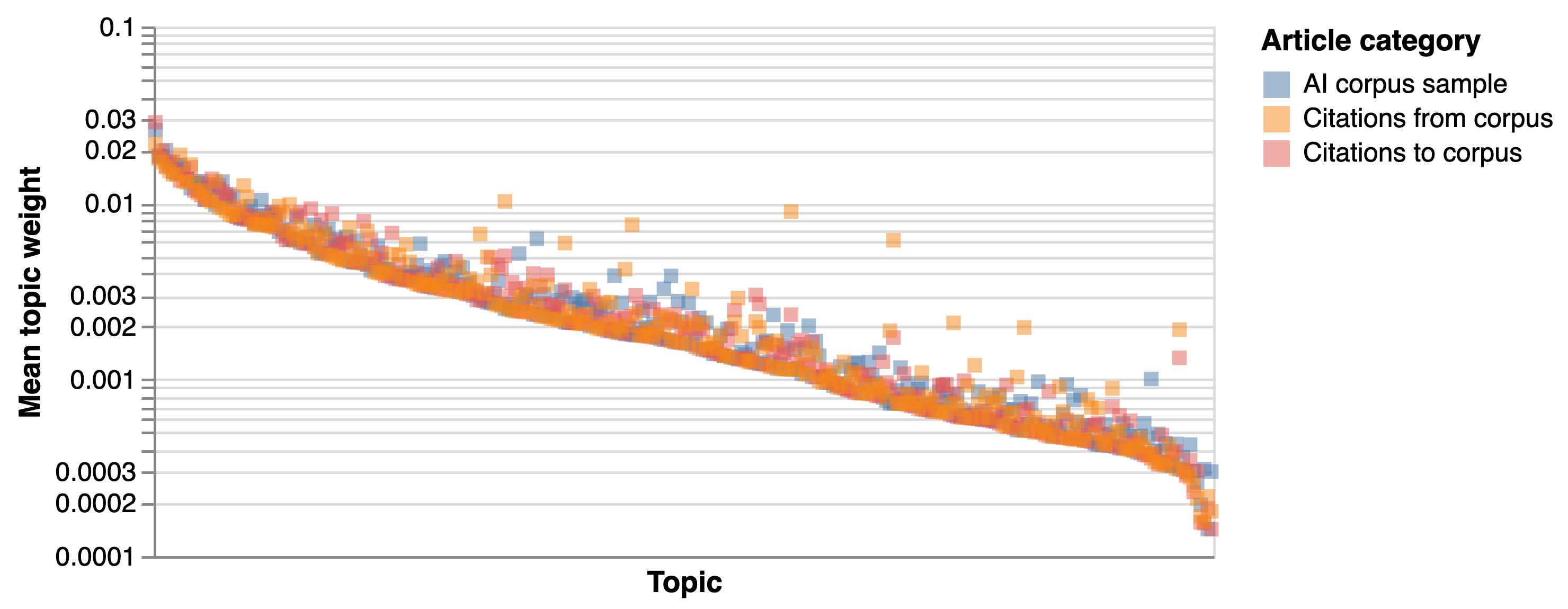}
    \captionsetup{margin=1cm}
    \caption{Average topic weight by article category, topics ranked from highest to lowest average weight in the horizontal scale with labels removed for readability.}
    \label{fig_rob:topic_comp}
\end{figure}

Figure \ref{fig_rob:topic_comp} shows the mean topic weight by sub-corpus, with topics ranked in the horizontal axis from highest to lowest mean weight in the corpus. The chart suggests strong similarities between the topic distributions of the papers sampled from the AI corpus and the papers citing to / from our corpus that are not part of it. The Spearman $\rho$ rank correlation between the mean topic weights in the AI sample and the citing from / citing to sub-corpora are respectively 0.96 and 0.92.

While the aggregated picture reveals strong similarities between the three distributions, figure \ref{fig_rob:topic_comp} shows some cases where topics are strongly represented in one of the sub-corpora but not others. Table \ref{tab:rob_over} presents the ten most over-represented topics in each sub-corpus calculated as the mean weight of the topic for a sub-corpus normalised by its average weight in all corpora. The \textit{Citations from AI corpus category} includes topics related to physics and material sciences, reflecting the fact that AI papers draw on other disciplines which we would not expect to see included in our AI corpus. Perhaps more worryingly, the \textit{Citations to AI corpus} category includes various topics related to computer vision and natural language processing suggesting that our corpus may be missing some applications of AI techniques in other domains. This leads to use an alternative operationalisation of AI also including citations in the robustness tests for our analysis of the evolution of thematic diversity.

\begin{table}[]
    \small
    \centering
\begin{tabular}{p{0.75in}>{\raggedright\arraybackslash}p{4.75in}}
\toprule
  Category & Ten most overrepresented topics \\
\midrule
        AI sample &  ontology\_ontologies\_semantic, algorithm\_optimization\_evolutionary, coherence\_granules\_gait, logic\_semantics\_reasoning, clinical\_medical\_patients, mle\_poems\_parameterization, knowledge\_rules\_reasoning, time\_series\_forecasting, behavior\_learning\_skills, customers\_product\_business \\ \\
 Citations from AI &                                        galaxies\_galaxy\_data, mass\_planets\_stars, model\_systems\_system, gravitational\_wave\_signals, evolution\_gas\_formation, quantum\_classical\_states, dynamics\_system\_systems, spectral\_energy\_spectrum, music\_physics\_jet, materials\_properties\_energy \\ \\
 Citations to AI &                             translation\_nmt\_language, compression\_coding\_image, robustness\_adversarial\_adversarial, object\_detection\_detection, images\_generated\_generator, gan\_samples\_subspace, twitter\_bots\_social, image\_images\_style, super\_resolution\_high, video\_temporal\_videos \\
\bottomrule
\end{tabular}
    \caption{Top ten most overrepresented topics in each validation sub-corpus}
    \label{tab:rob_over}
\end{table}

\subsection*{A network analysis of the composition of AI research}

\begin{figure}[h!]
    \centering
    \includegraphics[width=0.9\textwidth]{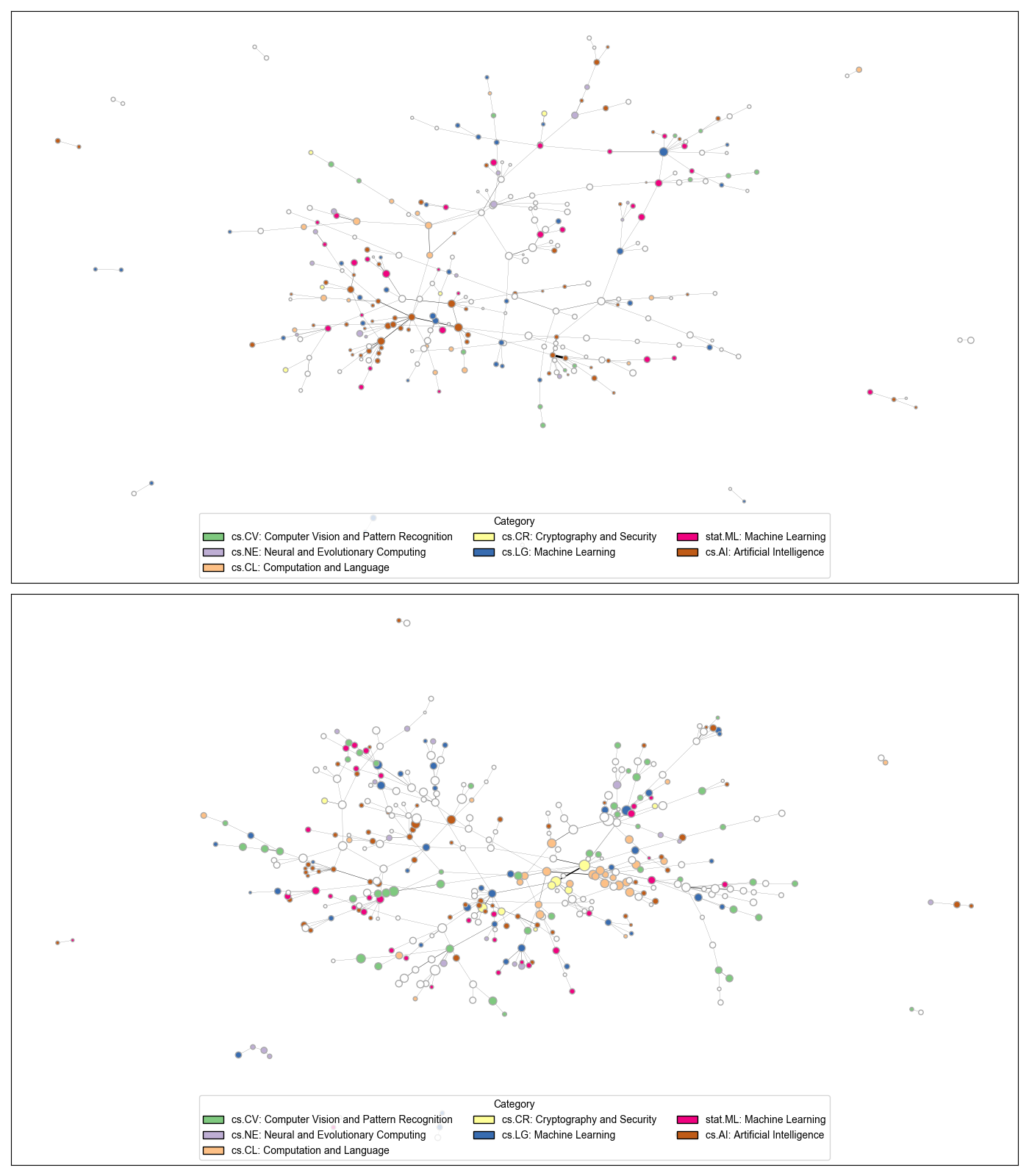}
    \captionsetup{margin=1cm}
    \caption{Topic co-occurrence graphs for 2013-2016 (top panel) and 2019-2020 (bottom panel). Each graph is the maximum spanning tree of the network of topic co-occurrences in articles published in the year above a minimum threshold of 0.1. The size of a node represents the number of instances its topic occurs in the topic and its colour the arXiv category where it is most salient.}
    \label{fig:topic_graph}
\end{figure}

We dig further into the evolution of the network structure of our corpus by comparing topic co-occurrence networks between 2013-2016 and 2019-2020. In these networks, each node is a topic and the edges between them are the number of times they co-occur in articles.\footnote{We consider that a topic occurs in an article if the probability estimated by our topic model is higher than 0.1}. We display the resulting networks in Figure \ref{fig:topic_graph}, where in order to simplify the visualisation we filter the co-occurrence network with a maximum spanning tree algorithm that preserves those edges with the largest weights that return a maximally connected network. The size of a node represent the topic's number of occurrences in the period, and its colour the arXiv category where the topic is most salient (calculated using the approach that we outlined in sub-section \ref{subsec:ai_evol}). Those topics that are not salient in any of the key categories in the legend are displayed with transparent labels. 

The networks illustrate the thematic transition undertaken by AI research in recent years. The 2013-2016 network displayed in the top panel is dominated by topics related to \textit{cs.AI} and \textit{stat.ML}. The 2019-2020 network in the bottom panel is more densely connected, and deep learning-related categories such as \textit{cs.CV}, \textit{cs.CL} and \textit{cs.CR} appear more prominently, in several cases forming communities of frequently co-occurring topics. The statistics of network connectivity and distance presented in table \ref{tab:network_metrics} are consistent with the idea that the distance between topics in the topic co-occurrence network have decreased over time in a way that could reduce disparity and lead to the stagnation or decline of diversity visible in Figure \ref{fig:div_evol}: the older network had more disconnected components, and, in its largest connected component, a longer diameter (maximum distance between topics), and a larger average path length (average number of steps that have to be taken to reach all other topics in the network from a given topic). 

\begin{table}[]
    \centering
    \begin{tabular}{lrrr}
    \toprule
        Network &  Number of components &  Average Path Length &  Diameter \\
        \midrule
        Network 2013-2016 &        13 &             5.823 &      14 \\
        Network 2019-2020 &        10 &             5.157 &      12 \\
        \bottomrule
        \end{tabular}
    \captionsetup{margin=1cm}
    \caption{Network statistics for topic co-occurrence networks including number of (disconnected) components in the network, average path length between topics in the largest component of the network and diameter (maximum distance between nodes in the largest connected component).}
    \label{tab:network_metrics}
\end{table}

\end{document}